\documentclass[12pt]{article}

\usepackage[margin=1in]{geometry}
\usepackage{amsmath,amssymb,amsthm}
\usepackage{booktabs,microtype}
\usepackage[authoryear,round]{natbib}
\usepackage[colorlinks=true,allcolors=blue]{hyperref}
\usepackage[nameinlink,noabbrev]{cleveref}

\newtheorem{theorem}{Theorem}[section]
\newtheorem{proposition}[theorem]{Proposition}
\newtheorem{corollary}[theorem]{Corollary}
\newtheorem{lemma}[theorem]{Lemma}

\theoremstyle{definition}
\newtheorem{definition}[theorem]{Definition}
\newtheorem{example}[theorem]{Example}

\theoremstyle{remark}
\newtheorem{remark}[theorem]{Remark}

\newcommand{\R}{\mathbb{R}}
\newcommand{\Sspace}{\mathbb{S}}
\newcommand{\E}{\mathbb{E}}
\newcommand{\Cov}{\operatorname{Cov}}
\newcommand{\Var}{\operatorname{Var}}
\newcommand{\col}{\operatorname{col}}
\newcommand{\rank}{\operatorname{rank}}
\newcommand{\range}{\operatorname{Range}}
\newcommand{\nullsp}{\operatorname{Null}}
\newcommand{\diag}{\operatorname{diag}}
\newcommand{\tr}{\operatorname{tr}}
\newcommand{\ip}[2]{\left\langle #1,#2\right\rangle_F}
\newcommand{\norm}[1]{\left\lVert #1\right\rVert_F}
\newcommand{\bu}{\hat u}
\newcommand{\Qclass}{\mathcal Q}
\newcommand{\Proj}{\Pi}

\title{Covariance Identification from Regression Residuals}

\author{Yanping Chen\thanks{Department of Economics, Indiana University Bloomington. Email: \texttt{chenyanp@iu.edu}. This paper is a theoretical companion to \citet{chen2026robust}, which develops estimation and inference for the sampling variance of the OLS estimator under the same projection geometry.}}

\date{\today}

\begin{document}

\maketitle

\begin{abstract}
Regression residuals are projections of the errors, not observations of them. This paper characterizes what residuals identify about the error covariance. Residualization acts on the covariance through a known linear map, and under an unrestricted covariance model the directions it destroys are exactly those generated by the column space of the regressors. A linear covariance functional is identified if and only if it does not load on these lost directions, and restricting the covariance model, for instance to diagonal or clustered form, shrinks them. The central result separates \emph{projection loss} from \emph{moment-selection loss}: projecting the residual covariance onto the maintained covariance structure loses no identifying information, so cross-cluster, long-lag, and long-distance residual products add no identifying directions. They can still improve conditioning, since restricting to the aligned moments squares the condition number. This equivalence relies on the orthogonal-projection geometry of OLS, and generally fails for non-orthogonal transformation. Quadratic statistics in the outcome that are free of the regression coefficients additionally identify the covariance between the fitted and residual components, but not the covariance within the span of the regressors. Because the OLS sandwich variance depends only on that block, it is not identified by any such statistic under an unrestricted covariance model; robust variance estimation draws its identification from maintained restrictions on the covariance structure.
\end{abstract}

\noindent\textbf{Keywords:}
regression residuals; covariance identification; orthogonal projection; sandwich variance; clustered dependence; HAC; network dependence.\\
\textbf{JEL codes:} C10, C18, C21, C23.

\section{Introduction}\label{sec:introduction}

Regression residuals are observable, but they are not observations of the disturbances. They are projections of the disturbances determined by the fitted design. This distinction underlies classical residual analysis for specification checking, outlier detection, and influence analysis \citep{anscombe1963examination, cook1982residuals}. More recently, work on heteroskedasticity-robust inference and variance components has emphasized that leverage can substantially distort the variance information contained in ordinary least-squares residuals, particularly when the number of covariates is large \citep{chesher1987bias, cattaneo2018inference, jochmans2022heteroscedasticity, kline2020leave}.

A natural response to this distortion is to construct an estimator of a particular covariance functional, such as the sampling variance of a regression coefficient. A separate and logically prior question is whether that functional is recoverable from the available regression-generated information in the first place. This paper draws that identification boundary exactly and characterizes both of its sides: which covariance directions regression destroys, and which functionals survive.

Conditional on a fixed design $X$, the population covariance of the OLS residual vector is
\begin{equation}\label{eq:operator-intro}
    \mathcal A(\Sigma) := M \Sigma M, \qquad M = I-X(X'X)^{-1}X'.
\end{equation}
Residualization therefore maps the latent error covariance $\Sigma$ into the population second moment of an observable through a known linear operator. Since the operator is generally non-injective, distinct latent covariance matrices can generate exactly the same population residual covariance. The relevant question is not merely the rank of $M$, but which directions in a maintained covariance model survive the transformation $\Sigma \mapsto M\Sigma M$.

Two clarifications about the word identification are needed at the outset, since everything below is stated at a fixed design and finite $n$. First, what gives the problem content is a restriction on admissible statistics, not on the data-generating process. The distribution of $y$ at the design determines $\E[y \mid X]$ and $\E[yy' \mid X]$, hence $\beta$ and $\Sigma$ with it, so in the usual sense nothing is unidentified and the null spaces computed here would be trivial. What is unavailable is the disturbance vector: the feasible object is the residual $\bu = My$, whose population second moment is $M \Sigma M$ rather than $\Sigma$. That substitution is an exact identity at every $n$, not an approximation, so the loss it produces is a matter of projection rather than of estimation error. Identification below therefore means recoverability from the population expectations of a restricted class of statistics: quadratic forms $y'Ay$ whose expectation is free of $\beta$, equivalently with $X'AX = 0$. This is the requirement imposed in restricted maximum likelihood, where inference on covariance parameters is based on error contrasts \citep{patterson1971recovery, harville1977maximum}, and the class is not one chosen to make the analysis work: HC and CR corrections, HAC, spatial and network estimators, and leave-out corrections are all quadratic forms of this kind, so characterizing what the class can and cannot deliver characterizes the reach of existing practice. \Cref{subsec:framework} develops the point.

Second, the non-identification results below do not contradict the asymptotic theory of robust inference. \citet{white1980heteroskedasticity} does not claim that the sandwich variance is recoverable at fixed $n$; the estimator is justified along a sampling sequence in which errors are independent across observations, and that independence is itself a restriction on the covariance model. \Cref{cor:sandwich-nonid} makes the role of the restriction explicit rather than disputing it. What does the identifying work is the maintained covariance space, and \cref{cor:sandwich-criterion} says how much structure that space must supply for a given contrast. The two analyses are complementary: the fixed-design statement holds at every $n$ and locates precisely which feature of the asymptotic framework is carrying the identification.

The paper distinguishes several nested information sets. At the narrowest level, an analyst may retain only selected or aggregated residual moments. The complete residual covariance contains weakly more information. Beyond residual moments lies the larger class of quadratic statistics in $y$ whose expectations are free of the regression coefficient. In terms of identification content, the hierarchy is
\begin{equation}\label{eq:information-hierarchy}
    P (M\Sigma M) \ \preceq\ M \Sigma M \ \preceq\ \left\{ \tr(A \Sigma): A = A',\ X' A X = 0 \right\},
\end{equation}
where $P(\cdot)$ denotes a structured collection of residual moments and $A \preceq B$ means that every linear covariance functional identified from $A$ is also identified from $B$. Each entry in \eqref{eq:information-hierarchy} is shorthand for the set of linear functionals it identifies, characterized in \cref{thm:functional-id,prop:invariance}.

The paper makes four contributions. First, it draws the boundary exactly and separates the two distinct mechanisms that produce it. Two admissible covariance matrices are observationally equivalent from residual second moments when their difference lies in the null space of $\mathcal A$, and under an unrestricted covariance model
\begin{equation}\label{eq:kernel-intro}
    \nullsp(\mathcal A) = \left\{ XB'+BX': B\in\R^{n\times p} \right\},
\end{equation}
so residualization discards precisely the symmetric covariance directions involving the column space of the design. On a restricted covariance space, identification depends on the intersection of this set with the maintained model, and a linear covariance functional $\ip{R}{\Sigma}$ is recoverable if and only if $R$ is orthogonal to every observationally invisible direction, equivalently if $R$ belongs to the range of the adjoint residual-information operator. This characterizes the functionals that survive residualization without requiring identification of the entire covariance matrix. It also isolates \emph{projection loss}, which occurs mechanically when $\Sigma$ becomes $M\Sigma M$, from \emph{moment-selection loss}, which occurs only afterward when the analyst retains a lower-dimensional collection of entries, blocks, or aggregates. Additional residual moments expand identification if and only if their inclusion shrinks the null space of the induced moment operator, which yields an exact rank criterion.

Second, the paper proves a sharp and somewhat counterintuitive equivalence theorem about moment-selection loss: more residual moment equations need not imply more identifying information. For any maintained linear covariance space $\mathcal H$, the orthogonal projection of $M\Sigma M$ back onto $\mathcal H$ has exactly the same null space as the complete residual covariance. Diagonal heteroskedasticity, known one-way clustering, unrestricted two-way covariance support, banded serial dependence, and known network support are leading examples. Residual products outside the maintained support may be nonzero and parameter-dependent, because projection redistributes dependence across observations, yet they contain no additional identifying directions conditional on the maintained model. Algebraically, the structure-aligned operator is the Gram operator associated with the full residual-information map, so the equivalence is a consequence of OLS residualization being an orthogonal projection. A symmetric invertible transformation that is not idempotent provides a counterexample, and so does an idempotent transformation that is not symmetric; the result is not generic to transformed disturbances.

Third, the same Gram representation shows that the equivalence is an identification statement and not a conditioning statement. Because $\mathcal A_S=\mathcal A^*\mathcal A$, the singular values of the structure-aligned operator are the squares of those of the full map, so $\kappa(\mathcal A_S)=\kappa(\mathcal A)^2$ on the identified subspace. The complete residual covariance is therefore exactly as informative as the aligned moments in population, and strictly better conditioned in finite samples whenever the full map is not already perfectly conditioned. This is the same squaring that motivates solving least-squares problems through the design matrix rather than the normal equations, and it explains why estimators built on aligned moments alone can behave poorly in designs where identification technically holds.

Fourth, the paper moves the boundary outward and shows where it stops. The mean-free class $\{A : X'AX = 0\}$ is strictly larger than the translation-invariant one generated by residual quadratic forms. It identifies all linear functionals of $M\Sigma M$ and of the cross block $H\Sigma M$, adding exactly $p(n-p)$ dimensions; leave-one-out moments of the form $\E[y_i\bu_i]$ operate in this margin and can identify heteroskedastic variance directions that residual second moments cannot. Only the $H\Sigma H$ component lies beyond every such statistic, and that is exactly the block on which the object of most applied interest depends. The sandwich variance of a linear contrast of $\widehat\beta$ has its representer inside $\nullsp(\mathcal A)$ under an unrestricted covariance model and depends on $\Sigma$ only through $H\Sigma H$, so it is identified neither by residual second moments nor by the larger mean-free class. Every heteroskedasticity- and cluster-robust procedure therefore obtains identification entirely from the maintained covariance space, and the paper gives dimension and leverage conditions describing what that space must deliver.

Two consequences for estimators in current use are worth flagging early, since they run in opposite directions. Under diagonal heteroskedasticity the residual squares are already information complete, so the many-covariate correction of \citet{cattaneo2018inference}, which inverts $M\circ M$ exactly, exhausts the residual-based information: no residual procedure can identify a variance direction it misses (\cref{cor:hadamard-complete}). Under two-way clustering the on-support residual entries are again information complete, but recovery is a different matter. The projection bias has no analogue that factors into per-cluster corrections, because the compression to the two-way support is not block diagonal, so there is no diagonal block to invert; the scalar adjustments in common use do not address the bias at all. Two-way support is where the moments an estimator reads and an exact inversion of the residual map stand furthest apart (\cref{subsec:twoway}).
\paragraph{Relation to the literature.}
The algebra of quadratic estimability has a long history in the statistical literature on variance components. \citet{seely1970linear,seely1971quadratic} study unbiased estimation over linear and quadratic spaces; \citet{rao1971estimation} develops MINQUE; and \citet{lamotte1973quadratic, horn1975estimating} characterize quadratic estimators of variance components. The present paper uses related linear-space geometry for a different purpose. Rather than beginning with a proposed estimator, it treats the population information generated by regression and residualization as the primitive and characterizes the resulting observational-equivalence classes of latent covariance matrices.

The distinction is particularly relevant for robust inference. Heteroskedasticity- and cluster-robust procedures replace latent error moments by functions of OLS residuals, while many-covariate corrections explicitly account for projection and leverage \citep{white1980heteroskedasticity, chesher1987bias, cattaneo2018inference, jochmans2022heteroscedasticity, kline2020leave}. Cluster, HAC, spatial, and network procedures additionally retain or aggregate residual products according to a maintained dependence structure \citep{cameron2011robust, bester2011inference, newey1986simple, andrews1991heteroskedasticity, conley1999gmm, kojevnikov2021limit}. The contribution here is to distinguish the identification questions underlying these practices: which covariance directions are destroyed by regression projection, which additional directions are lost through moment selection, and what information becomes available when one moves beyond residual-based quadratic statistics. The companion paper \citep{chen2026robust} takes the identified functional space characterized here as given and develops estimation and asymptotic inference for the sandwich variance within it.

These are population identification results, not efficiency results. Residual moments that are redundant for identification may nevertheless improve statistical precision, numerical stability, robustness to misspecification, specification testing, or structure learning; \cref{thm:conditioning} makes the precision channel exact. Conversely, when a target functional lies outside the identified functional space generated by a given information class, no estimator based solely on that class can recover it without additional restrictions.

The remainder of the paper is organized as follows.
\Cref{sec:setup} introduces the residual-information operator and characterizes its null space.
\Cref{sec:functionals} gives necessary and sufficient conditions for identification of linear covariance functionals.
\Cref{sec:selection} separates projection loss from moment-selection loss.
\Cref{sec:structure-aligned} establishes the structure-aligned equivalence theorem and the associated conditioning result.
\Cref{sec:invariance} enlarges the information set from residual second moments to all mean-free quadratic statistics.
\Cref{sec:identification} develops explicit identification and non-identification conditions and worked examples.
\Cref{sec:general-dependence} considers serial, spatial, and network covariance spaces.
\Cref{sec:discussion} concludes.

\section{Setup} \label{sec:setup}

Consider the fixed-design linear model
\begin{equation}\label{eq:model}
    y = X \beta + \varepsilon, \qquad \E[\varepsilon \mid X] = 0, \qquad \Cov( \varepsilon \mid X) = \Sigma,
\end{equation}
where $y,\varepsilon \in \R^n$, $X\in\R^{n\times p}$ has full column rank $p<n$. Let
\begin{equation}
    \Sspace^n := \left\{ A \in \R^{n\times n} : A = A' \right\}, \qquad
    \Sspace^n_+ := \left\{ A \in \Sspace^n : x'Ax\ge 0 \text{ for all } x \in \R^n \right\},
\end{equation}
and let $\Sigma \in \Sspace^n_+$. Throughout, statements are conditional on $X$. Let
\begin{equation}
    H = X(X'X)^{-1}X', \qquad M = I_n-H.
\end{equation}
Then the OLS residual vector is
\begin{equation}
    \bu  = My = M \varepsilon,
\end{equation}
and its population covariance is
\begin{equation}\label{eq:resid-cov}
    \Omega := \E[\bu\bu'\mid X] = M\Sigma M.
\end{equation}
Let $b_1,\ldots,b_n$ denote the standard basis vectors of $\R^n$, and let $x_i' = b_i'X$ denote the $i$th row of the design.

Identification is always relative to a model. Let $\mathcal H\subseteq\Sspace^n$ be a linear space containing all admissible covariance differences. Examples include the unrestricted symmetric matrix space, the diagonal space, a known block-diagonal cluster space, a banded serial covariance space, or a support space generated by a known spatial or network neighborhood relation.

Equip matrix spaces with the Frobenius inner product
\begin{equation}
    \ip{A}{B} := \tr(A'B),
\end{equation}
and write $\Proj_{\mathcal H}$ for the corresponding orthogonal projection of $\Sspace^n$ onto the linear subspace $\mathcal H$. For a support-restricted space, $\Proj_{\mathcal H}$ sets to zero the entries outside the support; for a Toeplitz space, it averages entries along equal lags.

Define the residual-information operator
\begin{equation}\label{eq:Adef}
    \mathcal A : \mathcal H\to\Sspace^n, \qquad \mathcal A(\Delta) = M\Delta M .
\end{equation}

Throughout, $\nullsp(\cdot)$ and $\range(\cdot)$ denote the null space and range of a linear operator.\footnote{We write $\nullsp(\cdot)$ rather than $\operatorname{Ker} (\cdot)$ because kernel functions appear in \cref{sec:general-dependence} in the HAC sense.}

\begin{definition}[Residual observational equivalence]
\label{def:equivalence}
Two admissible covariance matrices $\Sigma_0$ and $\Sigma_1$ are \emph{residual-second-moment equivalent} if
\begin{equation}
    M\Sigma_0 M = M \Sigma_1M,
\end{equation}
equivalently if $\Sigma_1-\Sigma_0 \in \nullsp(\mathcal A)$.
\end{definition}

\subsection{The meaning of identification} \label{subsec:framework}

This subsection clarifies the meaning of identification in this paper. In the usual sense, identification asks whether a population parameter is uniquely recoverable from the population distribution; here it asks whether it is uniquely recoverable from the population expectations of a restricted class of statistics.

At the fixed design $X$ the distribution of $y$ determines $\E[y \mid X]$ and $\E[yy' \mid X]$, hence
\begin{equation} \label{eq:trivial-id}
    \beta = (X'X)^{-1}X'\,\E[y \mid X], \qquad \Sigma = \E[yy' \mid X] - X\beta\beta'X' ,
\end{equation}
so $\Sigma$ is identified, and with it every functional of $\Sigma$, including the sandwich variance \eqref{eq:sandwich}. Read this way the null spaces computed below would all be trivial, so the content of the paper must lie elsewhere.

It lies in the step \eqref{eq:trivial-id} conceals: the formula recovers $\Sigma$ only because $\beta$ appears in it, and $\beta$ must in turn be replaced by something computable. Written in residual form the target is $\Sigma = \E[(y-X\beta)(y-X\beta)' \mid X]$; substituting the fitted value $Hy$ for the unknown mean $X\beta$ leaves the residual vector $\bu = My$, and
\begin{equation}\label{eq:projection-shift}
    \E[\bu\bu' \mid X] = M\Sigma M \ \neq\ \Sigma .
\end{equation}
The substitution involves no approximation and no sampling argument: \eqref{eq:projection-shift} is an exact identity at every $n$, and it is already where the information is lost. By \cref{prop:kernel-exact} the matrices $\Sigma$ and $\Sigma + XB' + BX'$ have the same image for every $B \in \R^{n\times p}$, so an analyst handed $M\Sigma M$ exactly, with no sampling error whatever, still could not tell them apart. What is missing is not data, and no estimator or bias correction supplies it. The source of the problem is the projection, not the estimation.

Equation \eqref{eq:projection-shift} is one instance of a general construction: any class of statistics induces a map from the covariance matrix to what that class can see in population. Fix a class $\mathcal C$ of symmetric matrices, each indexing a statistic $y'Ay$, and consider
\begin{equation} \label{eq:information-map}
    \Sigma \ \longmapsto\ \big\{\, \E[y'Ay \mid X] \ :\ A \in \mathcal C \,\big\} .
\end{equation}
\emph{Both sides of \eqref{eq:information-map} are population objects}: the left is the parameter, or a functional of it such as $\Var(c'\widehat\beta \mid X)$, the right what the statistics in $\mathcal C$ are aimed at. Identification in the usual sense compares $\Sigma$ against the whole distribution of $y$ and always succeeds, by \eqref{eq:trivial-id}; identification here compares it against \eqref{eq:information-map}, which is generally many-to-one. A covariance functional is \emph{identified} when it is constant on the fibres of \eqref{eq:information-map} --- \cref{thm:functional-id} turns this into an orthogonality condition --- and the null spaces computed below are the directions in $\Sigma$ the map collapses. Since the loss occurs in the map, it is repaired only by moving one of its two sides: restricting the maintained covariance space, as \cref{prop:monotone} describes, or enlarging $\mathcal C$, as \cref{sec:invariance} does. The remaining gap, between \eqref{eq:information-map} and the values $y'Ay$ actually computed, is estimation error; it shrinks along a design sequence and is the subject of \citet{chen2026robust}.

Three considerations select $\mathcal C$, and together they make the restriction a characterization rather than a convenience. The first fixes the shape of the statistic. The targets studied below are linear covariance functionals $\phi_R(\Sigma) = \ip{R}{\Sigma}$, so a usable statistic must have an expectation that depends on $\Sigma$, and depends on it linearly. Among functions of $y$ that property holds for every $\Sigma$ exactly on the quadratic forms $y'Ay$ with $A$ symmetric, since
\begin{equation}\label{eq:admissible}
    \E[y'Ay \mid X] = \beta'X'AX\beta + \ip{A}{\Sigma} .
\end{equation}
Relative to the class of targets, $\{y'Ay\}$ is therefore not one candidate set among several but the complete one.

The second is the difficulty that already produced \eqref{eq:projection-shift}, applied now to the statistic rather than to the target. Unless $X'AX = 0$, the expectation \eqref{eq:admissible} retains a term in $\beta$, so recovering the covariance part requires substituting $\widehat\beta$ and the error in $\widehat\beta$ contaminates what the statistic reveals about $\Sigma$ --- most severely when $p$ is large relative to $n$, the regime the many-covariate corrections of \citet{cattaneo2018inference} and \citet{kline2020leave} address. By \cref{lem:meanfree-invariant}(i) the expectation is free of $\beta$ if and only if $X'AX = 0$, and throughout an \emph{admissible statistic} is a quadratic form satisfying that condition. It is the requirement imposed in restricted maximum likelihood, where inference on covariance parameters is based on error contrasts \citep{patterson1971recovery, harville1977maximum}.

The third is that the resulting class is descriptive rather than formal. HC and CR corrections, HAC, spatial and network estimators, and leave-out corrections are all quadratic forms in $y$ with $X'AX = 0$, so a limit on what the class can deliver binds every robust variance estimator in current use at once. The scope of that claim should be read narrowly: it covers variance estimation, not inference generally. Procedures that avoid estimating the variance --- the bootstrap, or identification-robust confidence sets in the manner of Anderson--Rubin --- lie outside $\mathcal C$, and nothing below speaks against them. That is a feature rather than a gap, since \cref{cor:sandwich-nonid} is precisely a reason to consider them.

Within the admissible class, two nested subclasses are studied. The translation-invariant statistics $\{A : AX = 0\}$ are by \cref{lem:meanfree-invariant}(iii) exactly the quadratic forms in the residual vector, so their expectations are the linear functionals of $M\Sigma M$ --- the case of \eqref{eq:projection-shift}. This is the choice behind the operator $\mathcal A$ of \eqref{eq:Adef} and the maintained one until \cref{sec:invariance}, where the full mean-free class $\{A : X'AX = 0\}$ is shown to be strictly larger and to reach the cross block $H\Sigma M$ as well.

\begin{remark}[Relation to quadratic estimability]\label{rem:terminology}
The notion just defined is quadratic estimability in the sense of \citet{seely1970linear} and \citet{lamotte1973quadratic}, not identification in the sense of \citet{rothenberg1971identification}, which \eqref{eq:trivial-id} already delivers. Statements of the form ``the sandwich variance is not identified'' abbreviate the exact claim, that no $\beta$-free quadratic statistic has an expectation from which $\Var(c'\widehat\beta\mid X)$ can be recovered; the theorems below are stated in the exact form and should be read that way.
\end{remark}

\begin{remark}[Mean-freeness and sampling variance]\label{rem:variance-cost}
Statistics in the mean-free class that are not translation invariant are free of $\beta$ in expectation but not in distribution, which is the cost of leaving the residual class. Under Gaussian errors, $\Var(y'Ay\mid X) = 2\tr(A\Sigma A\Sigma) + 4\beta'X'A\Sigma A X\beta$, so the second term vanishes identically only when $AX=0$. For the leave-one-out statistic $y_i\bu_i$ of \cref{subsec:leaveout} it evaluates to
\begin{equation}\label{eq:leaveout-variance}
    (x_i'\beta)^2\,[M\Sigma M]_{ii},
\end{equation}
so the $p(n-p)$-dimensional enlargement of \cref{sec:invariance} is not free: it buys covariance directions at the cost of sampling behavior that is no longer invariant to the regression mean, and the cost grows with the fitted value at observation $i$.
\end{remark}

\subsection{Compression to the residual subspace}

Let $r = \rank(M) = n-p$. Choose $Q\in\R^{n\times r}$ with orthonormal columns spanning $\range(M)$.

\begin{proposition}[Compression to the residual subspace]
\label{prop:compression}
For every $\Delta\in\mathcal H$,
\begin{equation}\label{eq:compression}
    M\Delta M = Q(Q'\Delta Q)Q'.
\end{equation}
Consequently,
\begin{equation}\label{eq:kernel-compression}
    \nullsp(\mathcal A) = \left\{ \Delta\in\mathcal H: Q'\Delta Q=0 \right\}.
\end{equation}
Moreover, when $\mathcal H=\Sspace^n$ the map $\Delta\mapsto Q'\Delta Q$ is onto
$\Sspace^r$, so $\rank(\mathcal A)=r(r+1)/2$.
\end{proposition}

Thus residual second moments reveal only the compression $Q'\Sigma Q$, that is, the action of the latent covariance within the residual subspace.

Under an unrestricted covariance model, the discarded directions can be described explicitly.

\begin{proposition}[Exact null space under an unrestricted covariance model]
\label{prop:kernel-exact}
If $\mathcal H=\Sspace^n$, then
\begin{equation}\label{eq:kernel-exact}
    \nullsp(\mathcal A) = \left\{ XB'+BX': B\in\R^{n\times p} \right\},
\end{equation}
and
\begin{equation}
    \dim\nullsp(\mathcal A) = np-\frac{p(p-1)}{2}.
\end{equation}
\end{proposition}

For example, if $a\in\col(X)$, then $aa'\in\nullsp(\mathcal A)$. A covariance component loading entirely on the fitted-design space is therefore invisible to residual second moments. This observation is the seed of \cref{cor:sandwich-nonid} below.

For a restricted covariance space $\mathcal H$,
\begin{equation}\label{eq:restricted-kernel}
    \nullsp(\mathcal A) = \mathcal H \cap \left\{ XB'+BX': B\in\R^{n\times p} \right\}.
\end{equation}
Hence the entire covariance is identified from residual second moments exactly when $\nullsp(\mathcal A) = \{ 0 \}$.

Representation \eqref{eq:restricted-kernel} has an immediate monotonicity consequence that is easy to overlook: weakening the maintained covariance model never helps identification.

\begin{proposition}[Monotonicity in the maintained model]\label{prop:monotone}
Let $\mathcal H_1\subseteq\mathcal H_2\subseteq\Sspace^n$ be linear covariance spaces. Then
\begin{equation}
    \nullsp(\mathcal A|_{\mathcal H_1}) = \nullsp(\mathcal A|_{\mathcal H_2})\cap\mathcal H_1,
\end{equation}
and every linear functional identified on $\mathcal H_2$ is identified on
$\mathcal H_1$. The converse fails in general.
\end{proposition}

Robustness to the dependence structure and identification of covariance functionals are therefore in direct tension. Enlarging $\mathcal H$ to guard against misspecified dependence enlarges the null space and shrinks the identified functional space. The extreme case is $\mathcal H=\Sspace^n$, where \cref{cor:sandwich-nonid} shows the sandwich variance is lost entirely.

\subsection{The covariance cone}

The analysis uses a linear difference space $\mathcal H$, whereas covariance matrices must be positive semidefinite. The distinction is immaterial at interior parameter points.

\begin{lemma}[Cone-valid perturbations]\label{lem:cone}
Let $\Sigma_0\in\mathcal H$ be positive definite. For every $\Delta\in\mathcal H$, there exists $t>0$ such that
\begin{equation}
    \Sigma_0+s\Delta \in \mathcal H\cap\Sspace^n_+
\end{equation}
for all $|s|\leq t$.
\end{lemma}

Thus local observational equivalence at an interior covariance matrix is governed exactly by the linear null space in \eqref{eq:restricted-kernel}. At boundary points, one-sided restrictions can matter, but the linear characterization remains the relevant interior identification benchmark.

\section{Identifiable covariance functionals} \label{sec:functionals}

The entire covariance matrix is often not the parameter of interest. Let
\begin{equation}\label{eq:functional}
    \phi_R(\Sigma) := \ip{R}{\Sigma}, \qquad R\in\mathcal H,
\end{equation}
be a scalar linear covariance functional. There is no loss in taking $R\in\mathcal H$: a general symmetric $R$ acts on $\mathcal H$ only through $\Proj_{\mathcal H}R$.

Sandwich variances, variance components, covariance contrasts, and many dependence parameters can be represented in this form after conditioning on the regression design.

\begin{definition}[Residual identification of a covariance functional]
The functional $\phi_R$ is identified from residual second moments on $\mathcal H$ if it takes the same value at every pair of admissible covariance matrices that generate the same population residual covariance.
\end{definition}

\begin{theorem}[Identification of a linear covariance functional]
\label{thm:functional-id}
The following statements are equivalent:
\begin{enumerate}
    \item $\phi_R$ is identified from $M\Sigma M$ on $\mathcal H$;
    \item $\ip{R}{\Delta} = 0$ for all $\Delta\in\nullsp(\mathcal A)$;
    \item $R \in \nullsp(\mathcal A)^\perp$;
    \item $R \in \range(\mathcal A^*)$.
\end{enumerate}
\end{theorem}

The orthogonal complement is taken inside $\mathcal H$. The adjoint of the residual-information operator is
\begin{equation}\label{eq:adjoint-general}
    \mathcal A^*(D) = \Proj_{\mathcal H}(MDM), \qquad D\in\Sspace^n.
\end{equation}
Hence, if $R = \mathcal A^*(D)$, then
\begin{equation}\label{eq:representation}
    \ip{R}{\Sigma} = \ip{D}{M\Sigma M}.
\end{equation}

Equation \eqref{eq:representation} establishes existence of a residual-based representation of the target. It does not determine which representation should be used for estimation. Different $D$ matrices can have different sampling variance, conditioning, and robustness properties even though they identify the same population functional. Those are estimation questions rather than identification questions, and they are the subject of \citet{chen2026robust}.

\subsection{The sandwich variance}\label{subsec:sandwich}

The leading application concerns the sampling variance of a linear contrast of the OLS estimator. Conditional on $X$,
\begin{equation}\label{eq:sandwich}
    \Var(c'\widehat\beta\mid X) = c'(X'X)^{-1}X'\Sigma X(X'X)^{-1}c = \ip{R_c}{\Sigma},
\end{equation}
where
\begin{equation}\label{eq:Rc}
    R_c := X C X', \qquad C := (X'X)^{-1}cc'(X'X)^{-1}.
\end{equation}
The representer $R_c$ is thus supported entirely on the fitted-design space, which is exactly the region \cref{prop:kernel-exact} shows residualization destroys. Two consequences follow, and it is worth saying at the outset which one does the work. \Cref{cor:sandwich-nonid} is a benchmark: under an unrestricted covariance model the target is unavailable, and no choice of estimator repairs that. It is stated to locate where identification must come from, not because $\mathcal H=\Sspace^n$ is a model anyone maintains. The operative result is \cref{cor:sandwich-criterion}, which asks of a given maintained model and a given contrast how much of the destroyed region the target actually touches, and whose answer is often that it touches none of it.

\begin{corollary}[Non-identification of the sandwich variance]
\label{cor:sandwich-nonid}
Let $c\neq 0$ and let $\mathcal H=\Sspace^n$. Then:
\begin{enumerate}
    \item $R_c \in \nullsp(\mathcal A)$ and $\ip{R_c}{R_c}>0$, so $R_c\notin\nullsp(\mathcal A)^\perp$. Hence $\Var(c'\widehat\beta\mid X)$ is not identified from residual second moments.
    \item $\ip{R_c}{\Sigma}$ depends on $\Sigma$ only through $H\Sigma H$. Hence, by \cref{prop:invariance}, it is not identified by any mean-free quadratic statistic either.
\end{enumerate}
\end{corollary}

Under an unrestricted covariance model the sandwich variance is therefore not merely difficult to estimate; it is not a function of the population information generated by any $\beta$-free quadratic statistic, in the exact sense of \cref{rem:terminology}. Every heteroskedasticity- and cluster-robust procedure obtains identification from the maintained covariance space $\mathcal H$, not from the data. That statement is a starting point rather than a conclusion, because $\mathcal H=\Sspace^n$ is the extreme case and the practical question is how much structure a particular contrast requires. \Cref{thm:functional-id} converts it into a criterion that answers exactly that.

\begin{corollary}[Identification criterion for the sandwich variance] \label{cor:sandwich-criterion}
On a maintained covariance space $\mathcal H$, the functional $\Sigma \mapsto \Var (c' \widehat \beta \mid X)$ is identified from residual second moments if and only if
\begin{equation}\label{eq:sandwich-criterion}
    \ip{XCX'}{\Delta} = 0 \qquad\text{for every } \Delta\in\mathcal H\cap\{XB'+BX': B\in\R^{n\times p}\}.
\end{equation}
\end{corollary}

Condition \eqref{eq:sandwich-criterion} is easier to read after one substitution. Let
\begin{equation}\label{eq:influence}
    z := X(X'X)^{-1}c, \qquad\text{so that} \qquad c'\widehat\beta = z'y, \qquad R_c = zz', \qquad \Var(c'\widehat\beta\mid X) = z'\Sigma z .
\end{equation}
The vector $z$ collects the influence weights of the contrast, and the criterion becomes $z'\Delta z=0$ for every admissible invisible direction $\Delta$: what residualization hides must not load on the influence weights. Since $\ip{XCX'}{\Delta}=\tr(CX'\Delta X)$, a sufficient condition for \eqref{eq:sandwich-criterion} to hold simultaneously in every contrast is $X'\Delta X=0$ on $\mathcal H\cap\nullsp(\mathcal A)$. Residualization destroys covariance directions built from the column space of the design, so identification can fail only when the maintained dependence structure permits a covariance pattern the regressors could themselves have produced.

The condition looks restrictive, but it is satisfied in empirically common designs. The reason is that the covariance models used in practice are \emph{local}, permitting correlation only within single observations or small groups, whereas regressors are typically \emph{global}, varying across the whole sample. When the two do coincide, as under cluster fixed effects, the null space is never trivial and the criterion holds or fails according to which contrast is asked for. Two cases make the point.

\begin{example}[Diagonal heteroskedastic $\Sigma$]
\label{ex:diagonal-favorable}
Here $\mathcal H$ is the diagonal space, and an admissible difference $\Delta=\diag(\delta)$ has image
\begin{equation}\label{eq:diag-rankone}
    M\Delta M=\sum_{i=1}^n\delta_i\,m_im_i', \qquad m_i:=Mb_i .
\end{equation}
Consider first a nonnegative $\delta$. Then \eqref{eq:diag-rankone} is a sum of positive semidefinite terms, which vanishes only if each vanishes, so $\delta_i>0$ forces $m_i=0$, that is, $b_i\in\col(X)$: some combination of the columns of $X$ would have to be a point mass at observation $i$, equivalently that observation would have to have leverage one. Regressors that vary smoothly across the sample cannot single out one observation in this way.

A signed $\delta$ is also an admissible difference, by \cref{lem:cone}, and for it the argument fails: the terms in \eqref{eq:diag-rankone} may cancel even when none of them vanishes. What is required is that the $n$ matrices $\{m_im_i'\}$ be linearly independent, and their Gram matrix has entries $\ip{m_im_i'}{m_jm_j'} = (m_i'm_j)^2=M_{ij}^2$. The criterion is therefore nonsingularity of $M\circ M$, exactly as in \cref{prop:leverage}, and it holds in particular when no leverage exceeds one half. Residualization is what creates the possibility of cancellation: it replaces the point masses $b_i$, which are trivially independent, by the columns $m_i$ of $M$, which all lie in the $r$-dimensional residual subspace and may be nearly or exactly parallel; \cref{ex:matched-pairs} shows the gap is not vacuous. When $M\circ M$ is nonsingular the null space is trivial and both $X'\Delta X=0$ and $z'\Delta z=0$ hold with nothing to check.

Known clustering is the same argument with blocks in place of points. Write $W_g :=\{ w \in \R^n: w_i = 0 \text{ for } i \notin C_g\}$ for the vectors supported in cluster $g$, so $\dim W_g = n_g$. The rank-one terms $\delta_im_im_i'$ become $N_g \Delta_g N_g'$, where $N_g = ME_g$ collects the columns of $M$ indexed by $C_g$; a single-cluster difference $\Delta_g$ is invisible only if $N_g$ has deficient column rank, that is, only if $W_g\cap\col(X)\neq\{0\}$.\footnote{%
Write $E_g\in\R^{n\times n_g}$ for the selection matrix of cluster $g$, whose columns are $\{b_i:i\in C_g\}$, so that $v\mapsto E_gv$ is a bijection from $\R^{n_g}$ onto $W_g$ and $N_g = M E_g$. If $N_g v = 0$ for some $v\neq0$, then $w:=E_gv$ is a nonzero element of $W_g$ with $Mw = ME_gv = N_gv=0$, hence $w\in\nullsp(M)=\col(X)$ and $W_g\cap\col(X)\neq\{0\}$. Conversely, if $0\neq w\in W_g\cap\col(X)$, then $w=E_gv$ for a unique $v\neq0$ and $N_gv=Mw=0$, so $N_g$ has deficient column rank. The condition is therefore intrinsic to the pair $(W_g,\col(X))$ and does not depend on how the observations within $C_g$ are ordered. Note that for $n_g=1$ it reduces to $m_i=0$, since a single column has deficient rank only if it vanishes; for $n_g>1$ rank deficiency is strictly weaker than $N_g=0$.}
Cancellation across clusters is again what this condition misses: the null space is trivial only if the maps $\Delta_g\mapsto N_g\Delta_gN_g'$ are jointly injective on $\bigoplus_g \Sspace^{n_g}$, which by \cref{prop:dimension} requires $\sum_gn_g(n_g+1)/2\leq r(r+1)/2$. Vectorized, joint injectivity is full column rank of a Khatri--Rao product restricted to symmetric blocks, which reduces to nonsingularity of $M\circ M$ when every $n_g=1$; \cref{app:diagnostic} gives the matrix form. When it holds, the whole of $\Sigma$ is identified from residual second moments, not merely the sandwich variance and not merely for one contrast.
\end{example}

\begin{example}[Clustered $\Sigma$ with cluster fixed effects] \label{ex:fe-favorable}
Let $\mathbf 1_g\in\R^n$ denote the indicator vector of cluster $C_g$, with $(\mathbf 1_g)_i=1$ for $i\in C_g$ and $0$ otherwise, so that $\sum_g\mathbf 1_g=\mathbf 1_n$ and $\mathbf 1_g\in W_g$. Absorbing group effects means each $\mathbf 1_g$ is a column of $X$, so $\mathbf 1_g\in W_g\cap\col(X)$: exactly the coincidence the previous example ruled out, since a cluster dummy is a regressor confined to one cluster.

Since $M\mathbf 1_g=0$, every matrix $\mathbf 1_gw'+w\mathbf 1_g'$ with $w\in W_g$ satisfies $M(\mathbf 1_gw'+w\mathbf 1_g')M=(M\mathbf 1_g)w'M+Mw(M\mathbf 1_g)'=0$, and each is symmetric and supported inside block $g$, hence admissible. Therefore
\begin{equation}\label{eq:fe-kernel-family}
    \mathcal F := \operatorname{span} \left\{ \mathbf 1_gw'+w\mathbf 1_g' \ :\ g=1,\ldots,G,\ w\in W_g \right\} \ \subseteq\ \nullsp(\mathcal A|_{\mathcal H_C}) ,
\end{equation}
these being the elements of $\{XB'+BX'\}$ in \cref{prop:kernel-exact} obtained by taking $B$ supported on the dummy columns of $X$. The null space is therefore never trivial, and it is larger than the cluster levels alone: taking $w\neq\mathbf 1_g$ gives invisible directions carrying arbitrary within-cluster variation, not merely level shifts. When $X$ consists of the dummies alone, $\mathcal F$ is the whole null space and $\dim\mathcal F=\sum_gn_g$; adding further regressors enlarges $\col(X)$ and can enlarge the invisible directions beyond $\mathcal F$, as the design below illustrates.

The sufficient condition of \cref{cor:sandwich-criterion} fails here, since $X'\mathbf 1_g\mathbf 1_g'X=(X'\mathbf 1_g)(X'\mathbf 1_g)'\neq0$, so the full matrix $X'\Sigma X$ is not identified and no single argument covers every contrast at once. The criterion for an individual contrast is the weaker requirement $\ip{R_c}{\Delta}=z'\Delta z=0$, which can hold on the invisible directions even where $X'\Delta X\neq0$, and it does hold on $\mathcal F$ for any coefficient on a covariate varying within clusters. By Frisch--Waugh such a coefficient is estimated from the within-cluster residual $M_DX_1$, where $D=[\mathbf 1_1,\ldots,\mathbf 1_G]$ and $M_D=I-D(D'D)^{-1}D'$ demeans within each cluster separately, so the influence weights \eqref{eq:influence} satisfy $\mathbf 1_g'z=0$ for every $g$ individually. Every generator of $\mathcal F$ carries $\mathbf 1_g$ as a factor, so
\begin{equation}\label{eq:fe-orthogonality}
    z'\left( \mathbf 1_gw'+w\mathbf 1_g' \right)z = 2\,(z'\mathbf 1_g)(w'z) = 0 .
\end{equation}
What absorbing the group effects destroys is information about cluster \emph{levels}, on which a within-cluster coefficient does not depend; the invisible directions in $\mathcal F$ and the identifying direction are orthogonal by construction. A contrast on an absorbed effect is asking about a cluster level, which is exactly what the fixed effects removed.

A concrete design shows both the mechanism and the caveat about $\mathcal F$. Take $n=4$ with $C_1=\{1,2\}$, $C_2=\{3,4\}$ and
\begin{equation}
    X = [\,\mathbf 1_1,\ \mathbf 1_2,\ x\,] = \begin{pmatrix} 1&0&1\\ 1&0&-1\\ 0&1&1\\ 0&1&-1 \end{pmatrix}, \qquad x=(1,-1,1,-1)' .
\end{equation}
The columns are orthogonal with $X'X=\diag(2,2,4)$, so $r=1$, $\col(X)^\perp = \operatorname{span}\{q\}$ with $q=(1,-1,-1,1)'$, and
\begin{equation}
    M=\tfrac14qq', \qquad M\Sigma M=\tfrac1{16}(q'\Sigma q)\,qq', \qquad q'\Sigma q=\sum_{g=1}^2\left( a_g-2c_g+d_g \right),
\end{equation}
where $\Sigma_g=\left(\begin{smallmatrix}a_g&c_g\\c_g&d_g\end{smallmatrix}\right)$ denotes the cluster-$g$ block. The null space is the hyperplane $q'\Delta q=0$, so five of the six dimensions of $\mathcal H_C$ are lost, against the four of $\mathcal F$: the extra direction is $\Delta=\diag(1,0,-1,0)$, where neither cluster's contribution vanishes and the two offset. This is the cross-cluster cancellation of the previous example, unavoidable here since $\dim\mathcal H_C=6$ exceeds the bound $r(r+1)/2=1$ of \cref{prop:dimension}. For the contrast on $x$, that is $c=(0,0,1)'$, the influence weights are $z=X(X'X)^{-1}c=x/4$, and
\begin{equation}
    z'\Sigma z=\tfrac1{16}\,x'\Sigma x=\tfrac1{16}\,q'\Sigma q ,
\end{equation}
since within each cluster $x$ and $q$ are both $\pm(1,-1)$ and the sign is squared away: the target is a multiple of the one identified scalar. That coincidence is particular to this design: $x$ and $q$ have the same within-cluster amplitude in both clusters. Replacing $x$ by $(1,-1,2,-2)'$ gives $q = (2, -2, -1, 1)'$, and $\Delta = \diag(1,0,-4,0)$ is then invisible while $z' \Delta z \neq 0$, so the contrast variance is not identified even though $\mathcal{F}$ remains orthogonal to $z$. In general \eqref{eq:fe-orthogonality} disposes of $\mathcal F$, and the remainder of the null space must be checked against \cref{cor:sandwich-criterion} directly.

Two remarks complete the picture. First, within-cluster variation neither escapes the null space nor fills it. When $X$ consists of the dummies alone, $\mathcal F$ is the whole null space, of dimension $\sum_g n_g$ against $\sum_g n_g(n_g+1)/2$ for $\mathcal H_C$; what survives is the compression of each block to $\mathbf 1_{n_g}^\perp$, a fraction $(n_g-1)/(n_g+1)$ of the within-cluster structure. This is the best case rather than the typical one, since further regressors reduce $r$ and enlarge the invisible directions while leaving $\mathcal F$ unchanged: the design above loses five of six dimensions against the two of six the fraction would suggest. Second, tightening the maintained model can only help. Imposing idiosyncratic heteroskedasticity with a cluster random effect, or equicorrelation within clusters, shrinks $\mathcal H$ and by \cref{prop:monotone} shrinks the null space with it, so the criterion is preserved a fortiori. Under $\Sigma=\sigma^2I_n+\tau^2\sum_g\mathbf 1_g\mathbf 1_g'$, for instance, a within-cluster contrast has $z'\Sigma z=\sigma^2\|z\|^2$, identified while $\tau^2$ is not: the target never depended on what was lost.
\end{example}

\section{Projection loss and moment-selection loss}\label{sec:selection}

Residualization itself may destroy covariance information. Even after residualization, the analyst may discard further information by retaining only selected residual moments.

Let $\mathcal Y = \Sspace^n$ denote the complete residual-covariance space, and let $P: \mathcal Y\to\mathcal Z$ be a linear selection or aggregation operator into a finite-dimensional inner-product space $\mathcal Z$. Examples include extracting diagonal entries, retaining specified covariance blocks, averaging equal-lag entries, or reporting cluster-level summaries.

Define
\begin{equation}\label{eq:selected-op}
    \mathcal A_{P} := P\mathcal A.
\end{equation}
Since $\nullsp(\mathcal A) \subseteq \nullsp(\mathcal A_{P})$, moment selection can never create identifying information. It may or may not destroy information that survived residualization.

\begin{definition}[Information-complete residual moment selection]
\label{def:complete}
The selected moments are \emph{information complete} for $\mathcal H$ if
\begin{equation}\label{eq:kernel-equality}
    \nullsp(\mathcal A_{P}) = \nullsp(\mathcal A).
\end{equation}
\end{definition}

\begin{proposition}[Equivalent criteria for information completeness]
\label{prop:complete}
The following statements are equivalent:
\begin{enumerate}
    \item the selected residual moments are information complete;
    \item every covariance functional identified from the full residual covariance is also identified from the selected moments;
    \item $\range(\mathcal A_{P}^*) = \range(\mathcal A^*)$;
    \item $\rank(\mathcal A_{P}) = \rank(\mathcal A)$.
\end{enumerate}
\end{proposition}

The proposition formalizes a basic but important distinction: the number of observable moment equations is not the same as the number of identifying directions.

Full residual moments improve identification over selected moments exactly when there exists an admissible covariance perturbation $\Delta$ satisfying
\begin{equation}\label{eq:strict-loss}
    P\mathcal A(\Delta) = 0, \qquad \mathcal A(\Delta) \neq 0.
\end{equation}
For a particular functional $\phi_R$, the perturbation must additionally satisfy $\ip{R}{\Delta} \neq 0$.

\section{Structure-aligned residual information}\label{sec:structure-aligned}

A particularly natural selection is obtained by projecting the residual covariance back onto the maintained covariance space itself. Define the structure-aligned operator
\begin{equation}\label{eq:AS-general}
    \mathcal A_S: \mathcal H\to\mathcal H,
    \qquad
    \mathcal A_S(\Delta) = \Proj_{\mathcal H}(M\Delta M).
\end{equation}

For support-restricted covariance spaces, $\Proj_{\mathcal H}$ corresponds to retaining the residual entries allowed by the latent covariance support.\footnote{%
Under diagonal heteroskedasticity $\Proj_{\mathcal H_D}(W)=\diag(W_{11},\ldots,W_{nn})$, so alignment keeps the residual squares and discards every cross product. Under known clustering the same description holds in block form, $\Proj_{\mathcal H_C}(W)=\diag(W_{11},\ldots,W_{GG})$ with $W_{gg}$ the diagonal cluster blocks, keeping the within-cluster residual covariances; Step 1 of the proof of \cref{cor:cluster-equivalence} verifies that this coordinate description is indeed the orthogonal projection. In both cases the aligned moments are a subset of the residual entries.}
For other linear spaces, such as Toeplitz covariance models, it can involve aggregation rather than entry selection.\footnote{%
Under a stationary model $\mathcal H_{\mathrm{Toep}}$ the projection replaces each entry by the average of the entries at the same lag, so the aligned moments are the sample residual autocovariances that HAC procedures already form. Here no individual residual entry is retained: every aligned moment is a linear combination of all of them.}

\subsection{Gram representation}

\begin{lemma}[Adjoint and Gram representation]
\label{lem:gram-general}
For every $W\in\Sspace^n$,
\begin{equation}\label{eq:AF-adjoint}
    \mathcal A^*(W) = \Proj_{\mathcal H}(MWM).
\end{equation}
Moreover,
\begin{equation}\label{eq:AS-gram}
    \mathcal A_S = \mathcal A^*\mathcal A.
\end{equation}
Hence $\mathcal A_S$ is self-adjoint and positive semidefinite on $\mathcal H$.
\end{lemma}

Identity \eqref{eq:AS-gram} is the central algebraic observation of the paper. The structure-aligned residual-information operator is the Gram operator generated by the complete residual-information map. Everything in this section follows from that one identity together with the elementary fact that a linear map and its Gram operator have the same null space.

\begin{theorem}[Full--structure-aligned information equivalence]
\label{thm:general-equivalence}
For every linear covariance space $\mathcal H$ and every OLS residual-maker $M$,
\begin{equation}\label{eq:general-kernel-equality}
    \nullsp(\mathcal A_S) = \nullsp(\mathcal A).
\end{equation}
Consequently,
\begin{equation}\label{eq:general-range-equality}
    \range(\mathcal A_S) = \range(\mathcal A^*) \qquad \text{as subspaces of }\mathcal H.
\end{equation}
Therefore a linear covariance functional is identified from the complete residual covariance if and only if it is identified from the structure-aligned residual information.
\end{theorem}

The theorem does not say that excluded residual moments are constant, zero, or unrelated to the latent covariance. They can respond nontrivially to changes in $\Sigma$. The result says instead that, conditional on the maintained covariance space, those responses do not distinguish any pair of covariance matrices that cannot already be distinguished from the structure-aligned information.

This is the sense in which additional residual equations can be informative but redundant for identification.

\subsection{Why orthogonal projection matters} \label{subsec:counterexample}

The preceding theorem is not a generic property of transformed disturbances.

\begin{example}[Failure under a non-idempotent transformation] \label{ex:counterexample}
Let $n=2$ and let $\mathcal H$ be the diagonal covariance space. Consider
\begin{equation}
    T = \begin{pmatrix} 1&1\\ 1&-1 \end{pmatrix}, \qquad v = T\varepsilon.
\end{equation}
For $\Delta = \diag(\delta_1,\delta_2)$,
\begin{equation}
    T\Delta T' =
    \begin{pmatrix}
        \delta_1+\delta_2 & \delta_1-\delta_2 \\
        \delta_1-\delta_2 & \delta_1+\delta_2
    \end{pmatrix}.
\end{equation}
Taking $\delta_1=1$ and $\delta_2=-1$ makes both structure-aligned diagonal entries vanish while the off-diagonal entry is nonzero. Hence the aligned moments lose an identifying direction that remains visible in the full covariance of $v$.
\end{example}

The example isolates the role of OLS projection geometry. For a general transformation $T$, the Gram operator generated by $\Delta \mapsto T\Delta T'$ is
\begin{equation}
    \Delta \mapsto \Proj_{\mathcal H}\left( T'T\Delta T'T \right),
\end{equation}
which need not equal the structure-aligned transformation $\Delta \mapsto \Proj_{\mathcal H}(T\Delta T')$. The two agree precisely when
\begin{equation}\label{eq:TTcond}
    T'T = T,
\end{equation}
and \eqref{eq:TTcond} characterizes orthogonal projections: if $T'T=T$ then $T' = (T'T)' = T'T = T$, so $T$ is symmetric, and hence $T^2 = T'T = T$. Symmetry and idempotence are therefore not two hypotheses but the two halves of one, and the equivalence theorem is a statement about \emph{orthogonal} projection rather than about projection in general.

Each half is load-bearing. In \cref{ex:counterexample} the transformation is symmetric but not idempotent: $T'T = 2I_2 \neq T$, the Gram operator is $\Delta\mapsto4\Delta$, which is injective, and the aligned map nonetheless annihilates $\diag(1,-1)$. The next example shows that symmetry cannot be dropped either.

\begin{example}[Failure under an oblique projection] \label{ex:oblique}
Let $n=4$, let $\mathcal H$ again be the diagonal covariance space, and let
\begin{equation}
    T = \frac13
    \begin{pmatrix}
        3 & 2 & -1 & -1\\
        0 & 1 & 1 & 1\\
        0 & 2 & 2 & -1\\
        0 & 0 & 0 & 3
    \end{pmatrix}.
\end{equation}
Then $T^2=T$ and $\rank(T)=3$, so $T$ is a projection, but $T\neq T'$: it projects along a direction not orthogonal to its range. Take $\Delta = \diag(-1,3,-3,0)$. Every diagonal entry of $T\Delta T'$ vanishes, yet
\begin{equation}
    T\Delta T' = \frac13
    \begin{pmatrix}
        0 & 3 & 6 & 0\\
        3 & 0 & 0 & 0\\
        6 & 0 & 0 & 0\\
        0 & 0 & 0 & 0
    \end{pmatrix}
    \neq 0 .
\end{equation}
The aligned map has rank three on $\mathcal H$ while the full map has rank four, so a variance direction invisible to every diagonal moment remains visible in the off-diagonal moments. Idempotence alone is therefore not enough.
\end{example}

The mechanism is transparent in terms of the energy identity underlying \cref{thm:general-equivalence}. When $T'T=T$,
\begin{equation}\label{eq:energy}
    \norm{T\Delta T'}^2 = \ip{T\Delta T'}{\Delta},
\end{equation}
so all of the transformed covariance's energy is recovered by pairing it against $\Delta$. If $T\Delta T'$ were orthogonal to $\mathcal H$, then since $\Delta\in\mathcal H$ it would in particular be orthogonal to $\Delta$, and \eqref{eq:energy} would force $\norm{T\Delta T'}^2=0$. A nonzero image therefore cannot hide entirely in the moments that alignment discards. Both counterexamples exhibit exactly that configuration and so break \eqref{eq:energy}: in \cref{ex:counterexample} the left side is $8$ against a right side of $0$, and in \cref{ex:oblique} it is $10$ against $0$. The energy has migrated off the direction of $\Delta$ and into the discarded moments, which is what $T'T=T$ forbids.

\begin{remark}[Scope of the condition]\label{rem:scope-TT}
Condition \eqref{eq:TTcond} is sufficient but not necessary for a given $T$ and a given $\mathcal H$. Scalar multiples $T=cM$ with $c\neq0,1$ satisfy $T'T\neq T$ yet change no null space; and if $\mathcal H=\Sspace^n$ then $\Proj_{\mathcal H}$ is the identity, alignment discards nothing, and the equivalence holds trivially for every $T$. The correct reading is that \eqref{eq:TTcond} is what delivers the equivalence \emph{simultaneously for every linear covariance space}, which is the relevant guarantee because the analyst does not choose $\mathcal H$ to rescue the argument. Practically, the condition fails for GLS-weighted residuals, ridge-penalized fits, and any oblique partialling-out step, and the theorem should not be applied to them without verifying \eqref{eq:TTcond} directly.
\end{remark}

\subsection{Identification equivalence is not conditioning equivalence}
\label{subsec:conditioning}

\Cref{thm:general-equivalence} is a statement about null spaces. It says nothing about how well conditioned the two information sets are, and the Gram representation shows the difference is exactly a squaring.

\begin{theorem}[Conditioning of structure-aligned information]
\label{thm:conditioning}
Let $\sigma_1\geq\cdots\geq\sigma_m>0$ denote the nonzero singular values of $\mathcal A$ on $\mathcal H$, and let $\lambda_1\geq\cdots\geq\lambda_m>0$ denote the nonzero eigenvalues of $\mathcal A_S$. Then
\begin{equation}\label{eq:sv-squaring}
    \lambda_j = \sigma_j^2, \qquad j=1,\ldots,m,
\end{equation}
and hence, on the identified subspace $\nullsp(\mathcal A)^\perp$,
\begin{equation}\label{eq:cond-squaring}
    \kappa(\mathcal A_S) = \kappa(\mathcal A)^2, \qquad \kappa(\cdot):=\sigma_{\max}/\sigma_{\min}^{+}.
\end{equation}
In particular $\kappa(\mathcal A_S)\geq\kappa(\mathcal A)$, with equality if and only if $\kappa(\mathcal A)=1$.
\end{theorem}

The interpretation is worth stating plainly. Passing from the complete residual covariance to the structure-aligned moments is exactly the passage from a least-squares problem to its normal equations. The solution set is unchanged, which is \cref{thm:general-equivalence}; the conditioning is squared, which is \cref{thm:conditioning}. A covariance direction that residualization already preserves only weakly, so that $\sigma_j$ is small, produces a signal of size $\sigma_j$ in the full residual covariance and only $\sigma^2_j$ in the aligned moments. Since $\sigma_j \leq 1$, the aligned system is uniformly weaker, and disproportionately so exactly where the full map is already near failure.

Two practical implications follow. First, in designs where identification technically holds but $\mathcal A$ is poorly conditioned, estimators built exclusively on aligned moments can behave far worse in finite samples than estimators using the full residual covariance, even though the two are equivalent in population. The equivalence is pointwise in $n$ and therefore does not transfer automatically to a design sequence: an estimator can face an equivalent population problem at every sample size and still behave differently in the limit, a point taken up in \cref{subsec:varimp}. Second, numerical implementations should be organized around evaluations of $\mathcal A$ and $\mathcal A^*$ rather than around explicit formation of $\mathcal A_S$, for the same reason that iterative least-squares methods such as LSQR \citep{paige1982lsqr} are preferred to solving normal equations directly.

The equality case is not vacuous. \Cref{ex:cluster-fe-enlarged} below has $\kappa(\mathcal A)=1$, so the aligned and full information sets are equally well conditioned there, while \cref{ex:cluster-fe} has $\kappa(\mathcal A)=\sqrt3$ and the aligned system is strictly worse. Whether the squaring bites is a property of the design, not of the covariance model alone.

\begin{remark}[Where the excluded moments earn their place]
Taken together, \cref{thm:general-equivalence,thm:conditioning} give a precise accounting. Off-support residual products contribute nothing to identification under a correctly specified $\mathcal H$; they contribute to conditioning, hence to finite-sample precision and numerical stability; and, as \cref{subsec:overid} shows, they carry all the information available for testing whether $\mathcal H$ is correct in the first place.
\end{remark}

\subsection{Diagonal heteroskedasticity}\label{subsec:diagonal}

Let $\mathcal H_D = \{ \diag(\delta_1,\ldots,\delta_n): \delta\in\R^n \}$. For $\Delta = \diag(\delta)$,
\begin{equation}\label{eq:diag-diag}
    [M\Delta M]_{kk} = \sum_{i=1}^n M_{ki}^2\delta_i,
\end{equation}
so the structure-aligned operator is represented by the Hadamard square $M\circ M$. Cross-residual covariance moments satisfy
\begin{equation}\label{eq:diag-cross}
    [M\Delta M]_{k\ell} = \sum_{i=1}^n M_{ki}M_{\ell i}\delta_i.
\end{equation}
By \cref{thm:general-equivalence},
\begin{equation}\label{eq:diag-kernel}
    (M\circ M)\delta = 0 \quad\Longleftrightarrow\quad M\Delta M = 0.
\end{equation}
Thus cross-residual products cannot identify a heteroskedastic variance direction missed by every residual square.

\begin{corollary}[Information completeness of inverse-Hadamard corrections] \label{cor:hadamard-complete}
Under diagonal heteroskedasticity, the collection of squared-residual moments $\{\E[\bu_i^2]\}_{i=1}^n$ is information complete relative to the full residual covariance. Consequently, any procedure that inverts $M\circ M$ exactly, such as the many-covariate correction of \citet{cattaneo2018inference}, uses an information-complete moment set: no residual-based procedure can identify a variance direction it misses.
\end{corollary}

\subsection{Clustered dependence}\label{subsec:clusters}

Partition observations into nonempty clusters $C_1,\ldots,C_G$ of sizes $n_1,\ldots,n_G$. Suppose
\begin{equation}\label{eq:blockdiag}
    \Sigma = \diag(\Sigma_1,\ldots,\Sigma_G), \qquad \Sigma_g\in\Sspace^{n_g}_+ ,
\end{equation}
so that the admissible covariance space is
\begin{equation}\label{eq:HC}
    \mathcal H_C = \bigoplus_{g=1}^G \Sspace^{n_g}.
\end{equation}
This is the finite-sample form of the dependence structure maintained by cluster-robust variance estimation \citep{liang1986longitudinal,arellano1987computing}. Writing $M =(M_{gh})$ conformably,
\begin{equation}\label{eq:block-spillover}
    [M\Sigma M]_{gh} = \sum_{k=1}^G M_{gk}\Sigma_kM_{hk}' .
\end{equation}
Thus cross-cluster residual covariances can be nonzero even though the latent covariance is block diagonal. Let $P_C$ retain only diagonal cluster blocks.

\begin{corollary}[Known one-way cluster equivalence] \label{cor:cluster-equivalence}
Under \eqref{eq:blockdiag}, $\nullsp(P_C\mathcal A) = \nullsp(\mathcal A)$. Hence the complete collection of within-cluster residual covariance blocks identifies exactly the same linear covariance functionals as the full residual covariance.
\end{corollary}

\begin{corollary}[No identification rescue from cross-cluster residual products] \label{cor:no-cross-rescue}
There exists no $\Delta\in\mathcal H_C$ such that every within-cluster block of $M\Delta M$ is zero while at least one cross-cluster block is nonzero.
\end{corollary}

The statement is collective. One within-cluster residual product need not identify the corresponding latent covariance entry. Rather, the entire collection of within-cluster residual covariance blocks contains every identifying direction available in the full residual covariance.

It is also a statement about those blocks themselves, which is one compression short of what cluster-robust estimators report. They form a further $p\times p$ aggregate, $\Delta \mapsto \sum_g X_g' \Delta_{gg} X_g$, a selection operator in the sense of \cref{sec:selection} that is far from information complete relative to the whole latent covariance space: it compresses $\mathcal H_C$ onto a space of dimension at most $p(p+1)/2$. Yet it is lossless for the sandwich variance, since $\ip{R_c}{\Delta} = \tr\big(C\sum_g X_g' \Delta_{gg} X_g \big)$ depends on $\Delta$ only through the aggregate, so no direction the aggregate misses can move the target. Whether the target survives residualization at all remains the separate question settled by \cref{cor:sandwich-criterion}.

Identifying a functional and recovering it are different matters, however. Evaluated at the residual blocks as $\sum_gX_g'\bu_g\bu_g'X_g$, the aggregate has expectation $\sum_g X_g' [M\Sigma M]_{gg} X_g$ rather than the target $\sum_gX_g'\Sigma_gX_g$, since
\begin{equation}\label{eq:cluster-bias}
    \E[\bu_g\bu_g'\mid X] = M_{gg}\Sigma_gM_{gg}+\sum_{k\neq g}M_{gk}\Sigma_kM_{kg} .
\end{equation}
The corrections $\sum_gX_g'M_{gg}^{-\delta}\bu_g\bu_g'M_{gg}^{-\delta}X_g$ with $\delta=1/2$ or $1$ \citep{bell2002bias,imbens2016robust} address the first term of \eqref{eq:cluster-bias} block by block and leave the spillover untouched, so they are heuristic rather than exact inversions of $\mathcal A$; \cref{cor:hadamard-complete} records what exact inversion looks like in the diagonal case. \Cref{thm:functional-id} guarantees a residual representation exists whenever the target is identified, but not that it takes this form, and by \cref{thm:conditioning} it is better obtained through $\mathcal A$ and $\mathcal A^*$ than by inverting $\mathcal A_S$; which representation to use is an estimation question, taken up in \citet{chen2026robust}.

\subsection{Two-way support and the limits of local corrections}\label{subsec:twoway}

Two-way clustering separates the two questions further than any other case considered here, and in a direction that matters for practice: the moments a conventional estimator reads are information complete, while the bias correction it applies to them has no exact form.

The same argument covers two-way support. Suppose two observations may be correlated if they share either of two clustering dimensions, let $\mathcal S_{2W}$ contain all such
pairs, and define
\begin{equation}
    \mathcal H_{2W} = \left\{ \Delta\in\Sspace^n: \Delta_{ij}=0
      \text{ whenever } (i,j)\notin\mathcal S_{2W} \right\} .
\end{equation}
Since $\Proj_{\mathcal H_{2W}}$ again retains the entries in the support, \cref{thm:general-equivalence} applies verbatim: the residual covariance entries indexed by $\mathcal S_{2W}$ are information complete relative to the full residual covariance. What does not follow is that the conventional scalar cluster-score aggregates \citep{cameron2011robust, thompson2011simple} are. Such an aggregate reads only on-support entries, hence factors through the aligned moments, but it is a further selection applied to them and \cref{thm:general-equivalence} certifies only the alignment step; whether the aggregation preserves the null space is a separate rank question in the sense of \cref{prop:complete}.

For the sandwich variance the further compression is nonetheless lossless, by \cref{thm:functional-id}. Writing $\mathcal S_{2W}=\mathcal S_1\cup\mathcal S_2$ with $\mathcal S_{12}=\mathcal S_1\cap\mathcal S_2$, inclusion--exclusion gives
\begin{equation}\label{eq:two-way-inc-exc}
    \sum_{(i,j)\in\mathcal S_{2W}}x_ix_j'\Delta_{ij} = V_1(\Delta)+V_2(\Delta)-V_{12}(\Delta), \qquad V_S(\Delta):=\sum_{(i,j)\in S}x_ix_j'\Delta_{ij} ,
\end{equation}
so the three $p\times p$ aggregates the estimator assembles jointly determine $X'\Delta X$ on $\mathcal H_{2W}$, and the factorization $\ip{R_c}{\Delta}=\tr(CX'\Delta X)$ applies exactly as in the one-way case. The three terms are not a direct sum but a covering of $\mathcal S_{2W}$ corrected for overlap, which is what the subtraction records.

Recovery is where the two cases part company. Each aggregate is evaluated at $\bu_i\bu_j$ rather than $\Sigma_{ij}$, and $\E[\bu_i\bu_j\mid X]=[M\Sigma M]_{ij}$, but the bias \eqref{eq:cluster-bias} has no two-way analogue that factors into local corrections: the compression to $\mathcal S_{2W}$ is not block diagonal, so there is no $M_{gg}$ to invert and no per-cluster decomposition of the correction. The scalar adjustments in common use do not address the projection bias at all. Two-way support is therefore where the aligned moments and an exact inversion of $\mathcal A$ stand furthest apart, the diagonal and one-way estimators at least attacking the leading term of their own bias.

\begin{remark}[What is and is not settled for two-way clustering]\label{rem:twoway}
The accounting for $\mathcal H_{2W}$ has three separate layers, and conflating them is the source of the usual confusion. Identification: the on-support residual entries are information complete, by \cref{thm:general-equivalence}. Aggregation: the three $p\times p$ sums an estimator assembles determine $X'\Delta X$ on $\mathcal H_{2W}$ by \eqref{eq:two-way-inc-exc}, so the further compression is lossless for the sandwich variance even though it is far from information complete for $\Sigma$. Recovery: the aggregates are evaluated at $\bu_i\bu_j$, whose expectation is $[M\Sigma M]_{ij}$ and not $\Sigma_{ij}$, and here the case departs from the one-way one. Because the compression to $\mathcal S_{2W}$ is not block diagonal there is no $M_{gg}$ to invert and no per-cluster factorization of the correction, so the scalar degrees-of-freedom adjustments in common use address none of the projection bias. The diagonal case shows what the exact alternative looks like: invert $\mathcal A$ itself, as in \cref{cor:hadamard-complete}. No analogue is currently available on two-way support, which makes it the case where the aligned moments and an exact inversion of $\mathcal A$ stand furthest apart.
\end{remark}

\subsection{Overidentifying restrictions on excluded moments}\label{subsec:overid}

\Cref{thm:general-equivalence} has a converse reading that is useful for specification testing and, ultimately, for learning $\mathcal H$ itself.

\begin{proposition}[Excluded moments are pinned down]\label{prop:overid}
There exists a linear map $\mathcal T: \range(\mathcal A_S) \to \Sspace^n$ such that
\begin{equation}\label{eq:overid}
    (I-\Proj_{\mathcal H}) \big(M\Sigma M\big) = \mathcal T\big(\Proj_{\mathcal H}(M\Sigma M)\big) \qquad\text{for every } \Sigma\in\mathcal H.
\end{equation}
\end{proposition}

Under a correctly specified maintained model, the off-support residual moments are therefore a known linear function of the on-support ones. They contain no free parameters, which is exactly why they add no identifying directions; but their population values are restricted, which makes deviations from \eqref{eq:overid} testable. The number of overidentifying restrictions is $\dim\range(\mathcal A) - \dim\range(\mathcal A_S)$ measured in the ambient space, and these restrictions are the natural basis for testing a proposed cluster partition, support, or bandwidth. Population redundancy for identification and informativeness for specification testing are complementary, not contradictory.

\section{Beyond residual second moments}\label{sec:invariance}

The preceding sections condition on residual second moments as the information set. This section shows that this is itself a substantive restriction.

\subsection{Mean-free quadratic information}

For symmetric $A$,
\begin{equation}\label{eq:quad-expectation}
    \E[y'Ay\mid X] = \beta'X'AX\beta + \tr(A\Sigma).
\end{equation}
Define
\begin{equation}\label{eq:Qclass}
    \Qclass := \left\{ A\in\Sspace^n: X'AX=0 \right\}.
\end{equation}
Then $y'Ay$ has expectation depending on $\Sigma$ but not $\beta$ exactly when $A\in\Qclass$. The distinction between mean-freeness and residual invariance is important.

\begin{lemma}[Mean-freeness versus translation invariance]
\label{lem:meanfree-invariant}
For symmetric $A$:
\begin{enumerate}
    \item $\E[y'Ay\mid X]$ is free of $\beta$ if and only if $X'AX=0$;
    \item $(y+Xb)'A(y+Xb) = y'Ay$ for all $y,b$ if and only if $AX=0$;
    \item $AX=0$ if and only if $A=MAM$.
\end{enumerate}
\end{lemma}

The three parts chain together. By (ii) a quadratic form is unaffected by a shift in the regression mean exactly when $AX=0$, and by (iii) that condition is equivalent to $A=MAM$, so that
\begin{equation}
    y'Ay = y'MAMy = (My)'A(My) = \bu'A\bu :
\end{equation}
the translation-invariant statistics are precisely the quadratic forms in the residual vector, and their expectations are the linear functionals of $M\Sigma M$. By (i) the larger class asks only that $\beta$ drop out in expectation, and $AX=0$ implies $X'AX=0$, so residual quadratic forms constitute the translation-invariant subclass of the mean-free quadratic information set. \Cref{prop:invariance} measures the gap between them.

\subsection{The additional information margin}

\begin{proposition}[The mean-free quadratic information margin] \label{prop:invariance}
Let $r=n-p$. Then:
\begin{enumerate}
\item $\left\{ MDM: D\in\Sspace^n \right\} \subsetneq \Qclass$, with
\begin{equation}
    \dim\Qclass = \frac{n(n+1)}{2} - \frac{p(p+1)}{2}, \qquad \dim\left\{ MDM \right\} = \frac{r(r+1)}{2},
\end{equation}
so that
\begin{equation}\label{eq:invariance-gap}
    \dim\Qclass - \dim\{MDM\} = p(n-p).
\end{equation}

\item Every $A\in\Qclass$ decomposes as
\begin{equation}\label{eq:A-decomposition}
    A = MAM + HAM + MAH.
\end{equation}

\item Consequently, $\{ \tr(A\Sigma): A\in\Qclass \}$ identifies all linear functionals of $M\Sigma M$ and all linear functionals of $H\Sigma M$.

\item On $\mathcal H=\Sspace^n$, the null space of the map $\Sigma\mapsto\{\tr(A\Sigma):A\in\Qclass\}$ is exactly
\begin{equation}\label{eq:quad-kernel}
    \left\{ XCX' : C\in\Sspace^p \right\}, \qquad \dim = \frac{p(p+1)}{2}.
\end{equation}
Hence a linear functional depending only on the $H\Sigma H$ block is identified by the mean-free quadratic class if and only if it is the zero functional.
\end{enumerate}
\end{proposition}

The count in \eqref{eq:invariance-gap} is read most easily off \eqref{eq:A-decomposition}. Writing $A = MAM + HAM + MAH + HAH$, mean-freeness kills only the last block, of dimension $p(p+1)/2$, while translation invariance kills the cross blocks as well. Since $MAH = (HAM)'$, the free content of the pair is a single $p\times(n-p)$ matrix, and the margin is $p(n-p)$. Thus the covariance decomposes as
\begin{equation}
    \Sigma = M\Sigma M + M\Sigma H + H\Sigma M + H\Sigma H ,
\end{equation}
residual second moments identify the first block, mean-free quadratic statistics additionally identify the cross block, and the fitted-space block $H\Sigma H$ remains invisible, which is the content of \cref{cor:sandwich-nonid}. The enlargement is therefore not diffuse: it buys exactly one block and leaves the block of applied interest untouched.

\subsection{Leave-one-out moments}\label{subsec:leaveout}

The $p(n-p)$-dimensional enlargement has a concrete econometric interpretation. For observation $i$, define
\begin{equation}\label{eq:leaveout-A}
    A_i = \frac12 \left( b_ib_i'M + Mb_ib_i' \right).
\end{equation}
Then $A_i\in\Qclass$, since $MX=0$ annihilates both terms of $X'A_iX$, while $A_iX = \tfrac12(Mb_i)x_i'\neq0$ unless $Mb_i=0$: the leave-out matrix is mean-free but not translation invariant, and so is a witness to the strictness in \cref{prop:invariance}(i). Moreover $y'A_iy = y_i\bu_i$, so
\begin{equation}\label{eq:leaveout-moment}
    \E[y_i\bu_i\mid X] = (M\Sigma)_{ii},
\end{equation}
which under diagonal heteroskedasticity is $(1-h_{ii})\sigma_i^2$.

This is the population moment underlying leave-out corrections such as \citet{kline2020leave} and \citet{jochmans2022heteroscedasticity}. It involves the residual projection once rather than twice, which is why it can retain directions that disappear from $M\Sigma M$. Note that $A_i$ is not itself contained in the additional margin: its translation-invariant component $MA_iM = Mb_ib_i'M$ is generally nonzero. What matters is that $A_i$ has a nonzero component outside $\{MDM\}$, so that the moment is not a function of residual second moments alone. Nor do the $n$ matrices $\{A_i\}$ span the $p(n-p)$-dimensional margin; they occupy an $n$-dimensional slice of it. \Cref{rem:variance-cost} records the sampling cost of using them.

\begin{corollary}[Diagonal covariance identification under mean-free quadratic information] \label{cor:leaveout-id}
Suppose $\Sigma$ is diagonal. Then $\sigma_i^2$ is identified from the mean-free quadratic information class if and only if $h_{ii}<1$. Hence the full diagonal of $\Sigma$ is identified if and only if $\max_i h_{ii}<1$.
\end{corollary}

\subsection{A minimal strict-information-gain example}

\begin{example}[Matched pairs]
\label{ex:matched-pairs}
Let $n=2$, $p=1$, and $X=(1,1)'$. Then
\begin{equation}
    H = \frac12 \begin{pmatrix} 1&1\\ 1&1 \end{pmatrix}, \qquad M = \frac12 \begin{pmatrix} 1&-1\\ -1&1 \end{pmatrix} .
\end{equation}
Under diagonal heteroskedasticity,
\begin{equation}
    M\circ M = \frac14 \begin{pmatrix} 1&1\\ 1&1 \end{pmatrix},
\end{equation}
which has rank one, its null space spanned by $(1,-1)'$. Thus residual second moments identify only $\sigma_1^2+\sigma_2^2$: the two columns of $M$ satisfy $m_1=-m_2$, so the rank-one terms of \eqref{eq:diag-rankone} are proportional and cancel under a signed $\delta$, exactly the failure \cref{ex:diagonal-favorable} describes. Neither observation has unit leverage, however: $h_{11}=h_{22}=1/2$, and $\E[y_i\bu_i]=\tfrac12\sigma_i^2$, so each variance is separately identified under the larger mean-free quadratic information class. The second inequality in the information hierarchy \eqref{eq:information-hierarchy} is therefore strict in this design.
\end{example}

\section{Identification failures and their information content} \label{sec:identification}

The equivalence theorem determines whether selected residual moments lose information relative to the full residual covariance. It does not determine whether either collection identifies the covariance object of interest. A first constraint is a matter of counting.

\begin{proposition}[Residual-information dimension bound] \label{prop:dimension}
If $\nullsp(\mathcal A) = \{0\}$ on $\mathcal H$, then
\begin{equation}\label{eq:dim-bound}
    \dim(\mathcal H) \leq \frac{r(r+1)}{2}, \qquad r=n-p.
\end{equation}
\end{proposition}

The corresponding maximal dimension under the mean-free quadratic information class is $\tfrac{n(n+1)}{2} - \tfrac{p(p+1)}{2}$, which exceeds the residual-information dimension by $p(n-p)$. The bound is necessary and not sufficient, so a maintained model small enough to pass it may still be unidentified; the subsections below examine when.

\subsection{Diagonal heteroskedasticity and leverage}

\begin{proposition}[Leverage and diagonal covariance identification] \label{prop:leverage}
Suppose $\mathcal H$ is the diagonal covariance space.
\begin{enumerate}
    \item Residual second moments identify the full diagonal of $\Sigma$ if and only if $M\circ M$ is nonsingular.
    \item If $h_{ii}=1$ for some $i$, then $\sigma_i^2$ is not identified from residual second moments.
    \item If $\max_i h_{ii}<\tfrac12$, then $M\circ M$ is positive definite, and therefore the full diagonal of $\Sigma$ is identified from residual second moments.
\end{enumerate}
\end{proposition}

The two information classes can therefore be summarized as in \cref{tab:leverage}.

\begin{table}[t]
\centering
\caption{Identification of a diagonal covariance under the two information classes. The first and third rows are exact criteria; the second is merely sufficient. In particular $M\circ M$ can remain nonsingular even when some leverage exceeds $1/2$, so the gap between $1/2$ and $1$ should not be read as a comparison of necessary conditions.}
\label{tab:leverage}
\begin{tabular}{@{}p{0.58\textwidth}l@{}}
\toprule
    Information class & Condition for diagonal $\Sigma$\\
\midrule
    Residual second moments & $M\circ M$ nonsingular \\
    Residual second moments, simple sufficient condition & $\max_i h_{ii}<1/2$ \\
    All mean-free quadratic statistics & $\max_i h_{ii}<1$ \\
\bottomrule
\end{tabular}
\end{table}

\subsection{Two worked cluster designs}\label{subsec:worked}

Three points need illustrating: structure-aligned equivalence can coexist with substantial non-identification; moving beyond residual moments can enlarge the identified functional space; and even the larger information class need not identify a particular low-dimensional parameterization. No single design shows all three.

A design in which $X$ consists solely of cluster dummies is unsuitable for the first two, since there $M$ is block diagonal, every cross-cluster block of $M\Delta M$ vanishes identically, and \cref{cor:cluster-equivalence} holds trivially. \Cref{ex:cluster-fe} therefore includes a covariate varying within and across clusters, so that the cross-cluster residual covariances are nonzero and the equivalence has content; there the mean-free class recovers all of $\mathcal H_C$, which is why that design cannot carry the third point. \Cref{ex:cluster-fe-enlarged} returns to the cluster-dummy design, where the equivalence is degenerate and the mean-free gain is partial, but where a natural two-parameter submodel is unidentified under both classes. The two designs share $n=4$, the clusters $C_1=\{1,2\}$ and $C_2=\{3,4\}$, and the covariance space $\mathcal H_C = \Sspace^2 \oplus \Sspace^2$ with blocks $\Delta_g = \left( \begin{smallmatrix} a_g & c_g \\ c_g & d_g \end{smallmatrix} \right)$, so $\dim\mathcal H_C=6$; they differ only in the second column of $X$. \Cref{tab:example} collects the counts.

\begin{example}[Two clusters with a cluster-crossing covariate] \label{ex:cluster-fe}
Let
\begin{equation}
    X = \begin{pmatrix} 1&1 \\ 1&0 \\ 1&0 \\ 1&-1 \end{pmatrix},
\end{equation}
so that $p=2$ and $r=2$. The second regressor varies both within and between clusters, and the resulting $M$ is not block diagonal: direct computation gives $M_{13} = M_{24} = -1/4$ and $M_{14} = 1/4$, so cross-cluster residual covariances respond to the latent covariance. For $\Sigma=\diag(1,2,3,4)$, for instance, the cross-cluster block of $M\Sigma M$ has entries $(-3/4,\,5/8,\,-5/8,\,-1/2)$.

Since $r=2$, \cref{prop:dimension} already implies $\rank(\mathcal A)\leq3$, so at least three covariance directions are lost. Computing $\mathcal A$ on an orthonormal basis of $\mathcal H_C$ gives $\rank(\mathcal A)=3$ and $\dim \nullsp(\mathcal A)=3$, with nonzero singular values $(\sqrt3/2,\ \sqrt3/2,\ 1/2)$.

By \cref{thm:general-equivalence}, restricting attention to the two within-cluster residual blocks yields the same rank and the same identified functional space, even though the four cross-cluster residual entries are nonzero and vary with $\{\Delta_g\}$. The aligned operator's nonzero singular values are $(3/4,\ 3/4,\ 1/4)$, exactly the squares, so
\begin{equation}
    \kappa(\mathcal A)=\sqrt3, \qquad \kappa(\mathcal A_S)=3=\kappa(\mathcal A)^2,
\end{equation}
as \cref{thm:conditioning} requires. The two information sets agree exactly in identification content and differ in conditioning: the first point.

The mean-free quadratic class is strictly richer here, which is the second. Its null space on $\Sspace^4$ is $\{XCX':C\in\Sspace^2\}$, and a direct computation shows $\mathcal H_C\cap\{XCX'\}=\{0\}$: writing $X=[\mathbf 1,\ v]$ with $v=(1,0,0,-1)'$, the four cross-cluster entries of $XCX'$ are $C_{11}+C_{12}$, $C_{11}-C_{22}$, $C_{11}$, and $C_{11}-C_{12}$, which vanish only at $C=0$. The mean-free class therefore identifies all six dimensions of $\mathcal H_C$, against three for residual second moments. Structure-aligned equivalence and a strict gain from leaving the residual class coexist without difficulty.
\end{example}

That the mean-free class repairs the failure here might suggest it always does. It does not, and the cluster-dummy design shows why.

\begin{example}[Two clusters with cluster fixed effects] \label{ex:cluster-fe-enlarged}
Let $X$ contain the two cluster dummies instead. Then
\begin{equation}
    M = \diag(M_1,M_2), \qquad M_g = \frac12 \begin{pmatrix} 1&-1\\ -1&1 \end{pmatrix},
\end{equation}
which is block diagonal, so all cross-cluster residual covariances vanish identically and \cref{cor:cluster-equivalence} holds for the trivial reason. The residual covariance depends on cluster $g$ only through $a_g-2c_g+d_g$, so $\rank(\mathcal A)=2$ and $\dim\nullsp(\mathcal A)=4$. Here both nonzero singular values equal $1$, so $\kappa(\mathcal A)=\kappa(\mathcal A_S)=1$ and the squaring in \cref{thm:conditioning} is vacuous.

The mean-free quadratic class does better, though not as well as before. Writing the diagonal cluster blocks of a symmetric $A$ as $\left( \begin{smallmatrix} \alpha_g & \gamma_g \\ \gamma_g & \delta_g \end{smallmatrix} \right)$, the restriction $X'AX=0$ imposes three conditions: $\alpha_g + 2 \gamma_g + \delta_g = 0$ for $g=1,2$, together with $\mathbf 1_1'A\mathbf 1_2=0$ on the cross block. The third constrains only entries of $A$ that are annihilated when $A$ is paired with a block-diagonal $\Sigma$, so on $\mathcal H_C$ the operative restrictions are the first two, and
\begin{equation}
    \tr(A\Sigma) = \sum_{g=1}^2 \left( \alpha_ga_g + 2\gamma_gc_g + \delta_gd_g \right).
\end{equation}
The identified functional space on $\mathcal H_C$ therefore has dimension four rather than two, and the null space is $\mathcal H_C\cap\{XCX'\}=\operatorname{span}\{\mathbf 1_1\mathbf 1_1',\ \mathbf 1_2\mathbf 1_2'\}$, of dimension two rather than zero.

The third point follows once a submodel is imposed. Under the common equicorrelation model $a_g=d_g=\sigma^2$ and $c_g=\rho\sigma^2$, the space $\mathcal H$ has dimension two with tangent directions $T_\sigma = \partial \Sigma / \partial \sigma^2$ and $T_\rho = \partial \Sigma / \partial \rho$. Every residual-based identified functional depends on $(\sigma^2,\rho)$ only through $\sigma^2(1-\rho)$, and the same is true of the larger class, since the combination $T_\sigma+\tfrac{1-\rho}{\sigma^2}T_\rho$ equals $\mathbf 1_1\mathbf 1_1'+\mathbf 1_2\mathbf 1_2'$ and therefore lies in the mean-free null space. The identified dimension is one under both information classes, and $\sigma^2$ and $\rho$ are not separately identified by any $\beta$-free quadratic statistic.

The failure is not a consequence of clustering but of the fixed-effect design, which places the equicorrelation direction inside $\col(X)$. This is the same geometry that made \cref{ex:fe-favorable} the favorable case for the sandwich variance: fixed effects absorb covariance directions that are then simultaneously invisible and, for some targets, irrelevant.
\end{example}

\begin{table}[t]
\centering
\caption{
Identification in the two two-cluster designs. In both, the full and structure-aligned residual information sets have the same rank, as required by \cref{thm:general-equivalence}, and differ only in conditioning. The mean-free quadratic class is strictly richer in both, but repairs the equicorrelation failure in neither. }
\label{tab:example}
\begin{tabular}{@{}lccc@{}}
\toprule
& Full residual & Within-cluster & Mean-free \\
& covariance & blocks & quadratics \\
\midrule
\multicolumn{4}{@{}l}{
\emph{\Cref{ex:cluster-fe}: crossing covariate, $\dim\mathcal H_C=6$} } \\
    \quad Rank & 3 & 3 & 6 \\
    \quad Null-space dimension & 3 & 3 & 0 \\
    \quad Condition number & $\sqrt3$ & $3$ & --- \\
\addlinespace
\multicolumn{4}{@{}l}{
\emph{\Cref{ex:cluster-fe-enlarged}: cluster fixed effects, $\dim\mathcal H_C=6$} } \\
    \quad Rank & 2 & 2 & 4 \\
    \quad Null-space dimension & 4 & 4 & 2 \\
    \quad Condition number & $1$ & $1$ & --- \\
\addlinespace
\multicolumn{4}{@{}l}{
\emph{Equicorrelation submodel of \cref{ex:cluster-fe-enlarged}, $\dim\mathcal H=2$} } \\
    \quad Identified dimension & 1 & 1 & 1 \\
\bottomrule
\end{tabular}
\end{table}

\subsection{Implication for variance estimation}\label{subsec:varimp}

The identification results determine whether a population representation exists, not how it should be estimated.

If a target covariance functional lies in $\range(\mathcal A^*)$, then it admits a residual-based representation and variance estimation becomes an estimation problem involving the choice and statistical behavior of such a representation. Among the representations available, \cref{thm:conditioning} indicates that those built on the full residual covariance are better conditioned than those restricted to aligned moments, even though the two identify the same functional.

Whether that difference matters asymptotically turns on a single quantity, $\sigma_{\min}^+(\mathcal A)$. Along a design sequence on which it stays bounded away from zero the squaring in \cref{thm:conditioning} is a bounded factor; if it vanishes the two systems diverge, and disproportionately so in exactly the directions residualization already preserves most weakly. Establishing which case obtains requires a design sequence and a rate, neither of which the fixed-design analysis here supplies. \Citet{chen2026robust} studies the comparison under clustered dependence with many covariates and gives primitive leverage and cluster-size conditions. This is the point at which the present paper hands off.

If the target does not lie in $\range(\mathcal A^*)$, cross-cluster or other excluded residual products cannot repair the failure when structure-aligned equivalence applies. The target may nevertheless be identified by the larger mean-free quadratic information class, as in \cref{ex:matched-pairs}.

Only when the target is absent from the identified functional space of the larger class is the failure intrinsic to all $\beta$-free quadratic statistics. By \cref{cor:sandwich-nonid}, the sandwich variance under an unrestricted covariance model is exactly such a case, and the cluster equicorrelation submodel of \cref{ex:cluster-fe-enlarged} is another.

\section{Serial, spatial, and network dependence} \label{sec:general-dependence}

The structure-aligned equivalence theorem is not specific to clustering. It applies to any known linear covariance space.

\subsection{Serial dependence}

Suppose observations are ordered in time and latent covariance is known to vanish
beyond lag $L$:
\begin{equation}
    \mathcal H_L = \left\{ \Delta\in\Sspace^n: \Delta_{ts}=0 \text{ when } |t-s|>L \right\}.
\end{equation}
Then residual covariance entries with $|t-s|\leq L$ are structure aligned and contain every identifying direction present in the full residual covariance.

Residual products at longer lags may nevertheless be nonzero because residualization is global. Under the exact finite-range model, they add no identifying directions, though by \cref{prop:overid} their population values are restricted and hence testable.

If covariance stationarity is imposed instead, the maintained space may be Toeplitz:
\begin{equation}
    \mathcal H_{\mathrm{Toep}} = \left\{ \Delta: \Delta_{ts} = \gamma_{|t-s|} \right\}.
\end{equation}
Here $\Proj_{\mathcal H_{\mathrm{Toep}}}$ averages entries along equal lags, so the structure-aligned moments are precisely the sample residual autocovariances that HAC procedures already employ. This illustrates that structure alignment is a projection onto the maintained covariance space, not necessarily deletion of off-support matrix entries.

Kernel HAC methods such as those of \citet{newey1986simple} and \citet{andrews1991heteroskedasticity} generally impose approximate decay rather than exact finite-dimensional covariance restrictions, and the associated approximation and asymptotic issues are separate from the finite-sample comparison here. The comparison is nevertheless the fixed-$n$ shadow of what those arguments assume. A decay condition is a restriction on the covariance model, and by \cref{prop:monotone} restrictions can only shrink the null space; the bandwidth governs how much of $\mathcal H$ is surrendered in exchange for robustness to misspecified dependence, which is the same trade-off \cref{prop:monotone} describes in the exact case. The finite-range spaces above make that trade-off visible at every $n$ rather than only in the limit.

\subsection{Spatial and network dependence}

Let $d(i,j)$ be an observed distance measure. An exact finite-range covariance model is
\begin{equation}
    \mathcal H_d(c) = \left\{ \Delta: \Delta_{ij}=0 \text{ if } d(i,j)>c \right\}.
\end{equation}
Likewise, for an observed network $G_n$ with graph distance $d_G(i,j)$,
\begin{equation}
    \mathcal H_G(L) = \left\{ \Delta: \Delta_{ij}=0 \text{ if } d_G(i,j)>L \right\}.
\end{equation}
Under these exact support models, residual products within the maintained spatial or network neighborhood are information complete relative to the full residual covariance. Residual products outside the support may be generated by the regression projection but do not expand population identification. Spatial and network HAC procedures instead often allow covariance to decay gradually with distance \citep{conley1999gmm, kelejian2007hac, kojevnikov2021limit}. The exact support models above should therefore be interpreted as finite-dimensional benchmarks or sieves rather than as substitutes for the probabilistic assumptions needed for asymptotic HAC theory.

\subsection{Unknown and approximate covariance structure}

The equivalence result is conditional on a specified covariance space $\mathcal H$. If the support, cluster partition, network structure, or covariance model is unknown, the analyst cannot construct $\Proj_{\mathcal H}$ in advance.

In that setting, residual moments outside a candidate structure are the only available evidence about $\mathcal H$ itself. \Cref{prop:overid} makes this concrete: under a correct maintained model the off-support moments are a known linear function of the on-support ones, so departures identify misspecification. Redundancy for identification conditional on $\mathcal H$ does not imply redundancy for testing whether $\mathcal H$ is correct, and the moments that \cref{thm:general-equivalence} declares uninformative are precisely the ones that carry information about the structure.

For nonlinear covariance models $\Sigma(\theta)$ the global linear-space results do not apply directly, and the natural substitute is local. Fix $\theta_0$, write $T_k := \partial\Sigma(\theta)/\partial\theta_k\big|_{\theta_0}$, and let $\mathcal H_{\theta_0} := \operatorname{span}\{T_1,\ldots,T_K\}$ denote the tangent space. Every result above applies verbatim on $\mathcal H_{\theta_0}$, so $\theta$ is locally identified from residual second moments at $\theta_0$ whenever the images $\{\mathcal A(T_k)\}_{k=1}^K$ are linearly independent, equivalently whenever $\nullsp(\mathcal A|_{\mathcal H_{\theta_0}})=\{0\}$.

\begin{remark}[First-order and higher-order local identification]\label{rem:nonlinear}
Two cautions attach to reading that condition. First, it is a first-order condition, and its failure is not by itself a non-identification verdict. As in the parametric setting \citep{sargan1983identification, dovonon2013testing}, a parameter can be identified through second-order curvature while the tangent space is degenerate, so a rank deficiency among $\{\mathcal A(T_k)\}$ calls for a higher-order argument rather than a conclusion. Second, tangent spaces vary with $\theta$, so a constant-rank condition in the sense of \citet{rothenberg1971identification} is needed before local rank deficiency can be read as local non-identification at all. The equicorrelation submodel of \cref{ex:cluster-fe-enlarged} is a case where the negative conclusion does survive both cautions: there the offending combination $T_\sigma+\tfrac{1-\rho}{\sigma^2}T_\rho$ lies in the null space at every admissible $(\sigma^2,\rho)$, so the failure is uniform on the submodel rather than a first-order accident at one point, and no curvature argument can repair it. Spatial and network models with gradual decay are nonlinear in the same way and have not been checked against either caution here.
\end{remark}

\section{Discussion}\label{sec:discussion}

The paper yields four main lessons.

First, regression projection creates a fundamental covariance-identification boundary. Residual second moments reveal only $M\Sigma M$, or equivalently the compression of $\Sigma$ to the residual subspace. Under an unrestricted covariance model, the invisible directions are exactly $\{XB'+BX'\}$. Under a restricted covariance model, identification depends on which of these directions remain admissible. A covariance functional is recoverable precisely when it is orthogonal to the resulting null space.

Second, within the residual second-moment information class, more moment equations need not imply more identification. For any maintained linear covariance space, structure-aligned residual information is the Gram operator associated with the full residual-covariance map and therefore has exactly the same null space. Cross-cluster, long-lag, or long-distance residual products may respond to the latent covariance while remaining redundant for identification. This is a consequence of orthogonal projection --- of the condition $M'M=M$ in full, not of symmetry or idempotence alone --- rather than a generic feature of transformed errors.

Third, identification equivalence is not conditioning equivalence. The same Gram representation implies that the aligned operator has the squared singular values of the full map. The excluded moments buy no identifying directions and, except in the perfectly conditioned case, strictly better conditioning, which is why they can matter a great deal in finite samples and not at all in population.

Fourth, residual covariance is not the maximal regression-based covariance information set. Mean-free quadratic statistics satisfy $X'AX=0$, whereas translation-invariant residual quadratic forms satisfy the stronger condition $AX=0$. The difference adds $p(n-p)$ dimensions corresponding to the cross component $H \Sigma M$. Leave-one-out moments have a nonzero component in this additional margin and can therefore repair some identification failures that no residual second moment can repair. What no such statistic can repair is the loss of $H \Sigma H$, and that is exactly the block on which the sandwich variance depends. The reading of this is not that robust inference is unfounded but that its foundation is elsewhere: identification comes from the maintained covariance space, whether that space is imposed explicitly at fixed $n$ or implicitly through an asymptotic sampling sequence, and \cref{cor:sandwich-criterion} says what it has to deliver.

These results generate the hierarchy
\begin{equation}
    P(M\Sigma M) \ \preceq\ M\Sigma M \ \preceq\ \left\{ \tr(A\Sigma): X'AX=0 \right\}.
\end{equation}
Both inequalities can be strict, but the first becomes equality in identifying content for structure-aligned residual moments.

\subsection{Scope}

Several qualifications are important.

\emph{The results are finite-sample and conditional on $X$.} They do not contradict the asymptotic justification of robust inference, and consistency is not in question under the sequences for which it was established: the projection bias in \eqref{eq:projection-shift} vanishes with independent errors and $p$ fixed, which is why \citet{white1980heteroskedasticity} never required unbiasedness. What the fixed-design analysis adds is a criterion available at every $n$, and it binds hardest where those sequences are least credible --- many covariates, few clusters, and any maintained support that does not partition the observations, where the projection bias admits no per-block correction. More basically, consistency is defined relative to a sequence that itself supplies the covariance structure \cref{cor:sandwich-nonid} shows must come from somewhere other than the data; \cref{cor:sandwich-criterion} asks how much of it a given contrast actually requires, and \cref{ex:fe-favorable} shows the answer can be none at all.

\emph{Information equivalence is not covariance identification.} Two moment collections may have the same null space while both fail to identify many directions. The dimension bound and \cref{ex:cluster-fe} make this distinction explicit.

\emph{Identification equivalence is not statistical efficiency.}
Different representations of the same identified covariance functional can have different variance, conditioning, finite-sample bias, or robustness under misspecification. \Cref{thm:conditioning} quantifies one channel; the remainder is left to \citet{chen2026robust}.

\emph{The structure-aligned theorem depends on orthogonal-projection geometry.} The proof uses $M'M=M$, which characterizes orthogonal projections. Neither half can be dropped: \cref{ex:counterexample} exhibits a symmetric, invertible, well-conditioned transformation for which the equivalence fails, and \cref{ex:oblique} an idempotent but oblique one. The theorem should therefore not be applied to GLS-weighted residuals, penalized fits, or oblique partialling-out steps without verifying $M'M=M$ directly.

\emph{The theorem is conditional on the maintained covariance space.}
If the covariance support is misspecified, if cross-cluster dependence is allowed, or if the reported moment aggregates are not the orthogonal projection onto $\mathcal H$, excluded moments may add identifying information.

\emph{Population redundancy does not imply empirical irrelevance.}
Excluded moments can be useful for testing, structure learning, or robustness analysis even when they add no identifying direction under the maintained model.

\subsection{Extensions}

Three extensions follow naturally.

First, exact null-space comparisons can be replaced by spectral measures of near-identification. \Cref{thm:conditioning} gives the exact relationship between the two spectra, and small eigenvalues of $\mathcal A_S$ quantify covariance directions that are technically identified but only weakly preserved by residualization. For a specific target the natural index is
\begin{equation}\label{eq:near-id-index}
    \nu(R) := \frac{\big\|\Proj_{\nullsp(\mathcal A)^\perp} R\big\|_F}
                   {\big\|(\mathcal A^*)^{+} R\big\|_F},
\end{equation}
the ratio of the identified component of the representer to the norm of the residual representation that delivers it, with $(\mathcal A^*)^{+}$ the Moore--Penrose inverse on $\range(\mathcal A^*)$. Then $\nu(R)\geq\sigma^+_{\min}(\mathcal A)$ always, with equality when $R$ aligns with the weakest identified direction, and $\nu(R)\to0$ exactly when the target is identified only through directions residualization nearly destroys. A direction identified at every $n$ whose singular value shrinks along a design sequence is the covariance analogue of weak identification, and the rate at which $\nu$ shrinks is what separates a bounded conditioning penalty from an asymptotically consequential one. Establishing the connection to the sampling behavior of a particular estimator requires the design-sequence machinery this paper does not supply.

Second, the larger mean-free quadratic information class can be characterized jointly with the maintained covariance space. Comparing $\Proj_{\mathcal H}(\Qclass)$ with $\range(\mathcal A^*)$ would provide a general criterion for when leave-out-type information strictly enlarges the residual-based identified functional space.

Third, \cref{prop:overid} suggests treating $\mathcal H$ itself as the object of inference. Given a family of candidate covariance spaces, the overidentifying restrictions on excluded moments generate both specification tests and, potentially, a selection criterion. Characterizing when two candidate structures are observationally equivalent given $M\Sigma M$ is the natural next question, and it is the subject of ongoing work.

\newpage
\bibliographystyle{plainnat}
\bibliography{reference}

\newpage
\appendix

\section{Proofs}\label{app:proofs}

All matrix spaces below are finite dimensional and equipped with the Frobenius inner product $\ip{A}{B}=\tr(A'B)$; orthogonal complements and adjoints refer to this inner product throughout. Recall that $H=X(X'X)^{-1}X'$ and $M=I_n-H$ satisfy
\begin{equation}\label{eq:app-basic}
    M'=M,\quad M^2=M,\quad MX=0,\quad X'M=0,\quad H+M=I_n,\quad HM=MH=0 .
\end{equation}
These identities are used repeatedly without further comment.

\subsection{\texorpdfstring{Proof of \cref{prop:compression}}{Proof of compression}}

\begin{proof}
\emph{Step 1: $M=QQ'$.} By \eqref{eq:app-basic}, $M$ is symmetric and idempotent, hence the orthogonal projector onto $\range(M)$. Let $r=\rank(M)$ and let $Q\in\R^{n\times r}$ have orthonormal columns spanning $\range(M)$, so $Q'Q=I_r$. The matrix $QQ'$ is symmetric, satisfies $(QQ')^2=Q(Q'Q)Q'=QQ'$, and has range $\range(Q)=\range(M)$; it is therefore also the orthogonal projector onto $\range(M)$. Since the orthogonal projector onto a given subspace is unique,
\begin{equation}\label{eq:app-MQQ}
    M = QQ'.
\end{equation}

\emph{Step 2: the compression identity.} Substituting \eqref{eq:app-MQQ} twice and using associativity,
\begin{equation}
    M\Delta M=(QQ')\Delta(QQ')=Q\big(Q'\Delta Q\big)Q',
\end{equation}
which is \eqref{eq:compression}.

\emph{Step 3: the null space.} If $Q'\Delta Q=0$ then Step 2 gives $M\Delta M=Q0Q'=0$. Conversely, suppose $M\Delta M=0$. Premultiplying by $Q'$ and postmultiplying by $Q$ and using Step 2,
\begin{equation}
    0=Q'(M\Delta M)Q=(Q'Q)\big(Q'\Delta Q\big)(Q'Q)=I_r\big(Q'\Delta Q\big)I_r =Q'\Delta Q .
\end{equation}
Hence $M\Delta M=0$ if and only if $Q'\Delta Q=0$, which is
\eqref{eq:kernel-compression}.

\emph{Step 4: the rank when $\mathcal H=\Sspace^n$.} The map $\varphi:\Delta\mapsto Q'\Delta Q$ sends $\Sspace^n$ into $\Sspace^r$. It is surjective: given $S\in\Sspace^r$, the matrix $\Delta=QSQ'$ is symmetric and satisfies $\varphi(\Delta)=Q'QSQ'Q=S$. The map $\psi:S\mapsto QSQ'$ from $\Sspace^r$ to $\Sspace^n$ is injective, since $QSQ'=0$ implies $S=Q'(QSQ')Q=0$. As $\mathcal A=\psi\circ\varphi$ by Step 2 and $\varphi$ is onto $\Sspace^r$,
\begin{equation}
    \rank(\mathcal A)=\dim\psi(\Sspace^r)=\dim\Sspace^r=\frac{r(r+1)}{2}. \qedhere
\end{equation}
\end{proof}

\subsection{\texorpdfstring{Proof of \cref{prop:kernel-exact}}{Proof of exact null space}}

\begin{proof}
Write $\mathcal K:=\{XB'+BX':B\in\R^{n\times p}\}$, a linear subspace of $\Sspace^n$.

\emph{Step 1: $\mathcal K\subseteq\nullsp(\mathcal A)$.} For any $B\in\R^{n\times p}$, using $MX=0$ and $X'M=0$,
\begin{equation}
    M(XB'+BX')M=(MX)B'M+MB(X'M)=0+0=0 .
\end{equation}

\emph{Step 2: $\nullsp(\mathcal A)\subseteq\mathcal K$.} Let $\Delta\in\Sspace^n$ with $M\Delta M=0$. Inserting $I_n=H+M$ on both sides and expanding,
\begin{equation}\label{eq:app-expand}
    \Delta=(H+M)\Delta(H+M)=H\Delta H+H\Delta M+M\Delta H+M\Delta M =H\Delta H+H\Delta M+M\Delta H,
\end{equation}
the last equality by hypothesis. We express each surviving term as an element of
$\mathcal K$.

For the first term set $C:=(X'X)^{-1}X'\Delta X(X'X)^{-1}$, which is symmetric because $\Delta$ is. Then $H\Delta H=XCX'$, and with $B_1:=\tfrac12XC$,
\begin{equation}
    XB_1'+B_1X'=\tfrac12XC'X'+\tfrac12XCX'=XCX'=H\Delta H,
\end{equation}
using $C'=C$.

For the remaining two terms set $B_2:=M\Delta X(X'X)^{-1}$. Then, using $M'=M$ and $\Delta'=\Delta$,
\begin{equation}
    XB_2'=X(X'X)^{-1}X'\Delta M=H\Delta M,
    \qquad
    B_2X'=M\Delta X(X'X)^{-1}X'=M\Delta H .
\end{equation}
Setting $B:=B_1+B_2$ and adding the three displays gives $\Delta=XB'+BX'\in\mathcal K$.

\emph{Step 3: the dimension.} By \cref{prop:compression} with $\mathcal H=\Sspace^n$ we have $\rank(\mathcal A)=r(r+1)/2$ with $r=n-p$. Rank--nullity on $\Sspace^n$ gives
\begin{equation}
    \dim\nullsp(\mathcal A) =\frac{n(n+1)}{2}-\frac{(n-p)(n-p+1)}{2} =\frac{2np-p^2+p}{2} =np-\frac{p(p-1)}{2},
\end{equation}
where the middle equality expands $(n-p)(n-p+1)=n^2+n-2np+p^2-p$.
\end{proof}

\begin{remark}[Consistency of the dimension count]
The same number arises from the parameterization directly. The linear map $B\mapsto XB'+BX'$ from $\R^{n\times p}$ onto $\mathcal K$ has null space $\{B:XB'+BX'=0\}$. Writing $B=XC_1+MC_2$ with $C_1\in\R^{p\times p}$ and $MC_2$ the component orthogonal to $\col(X)$, the condition forces $C_2=0$ and $C_1'=-C_1$, so the null space is $\{XC:C'=-C\}$, of dimension $p(p-1)/2$. Hence $\dim\mathcal K=np-p(p-1)/2$, in agreement with \cref{prop:kernel-exact}.
\end{remark}

\subsection{\texorpdfstring{Proof of \cref{prop:monotone}}{Proof of monotone}}

\begin{proof}
\emph{Step 1: the null-space identity.} By \eqref{eq:restricted-kernel} applied to each space, $\nullsp(\mathcal A|_{\mathcal H_i})=\mathcal H_i\cap\mathcal K$ for $i=1,2$, where $\mathcal K=\{XB'+BX'\}$ as above. Since $\mathcal H_1\subseteq\mathcal H_2$,
\begin{equation}
    \mathcal H_1\cap\mathcal K = \mathcal H_1 \cap(\mathcal H_2\cap\mathcal K) = \nullsp( \mathcal A|_{\mathcal H_2} ) \cap \mathcal H_1 .
\end{equation}

\emph{Step 2: transfer of identification.} Let $R\in\mathcal H_1$ and suppose $\phi_R$ is identified on $\mathcal H_2$. By \cref{thm:functional-id}, $\ip{R}{\Delta}=0$ for every $\Delta\in\nullsp(\mathcal A|_{\mathcal H_2})$. By Step 1 the null space on $\mathcal H_1$ is a subset of that null space, so the same orthogonality holds there, and \cref{thm:functional-id} applied on $\mathcal H_1$ gives identification.

\emph{Step 3: the converse fails.} Let $\mathcal H_1=\mathcal H_D$ be the diagonal space, let $\mathcal H_2 = \Sspace^n$, and consider the total-variance functional $\phi_{I_n}(\Sigma) = \tr(\Sigma)$, whose representer is $R=I_n$ on either space.

On $\mathcal H_2$ the functional is not identified. Taking $B=X$ in \cref{prop:kernel-exact} gives $\Delta:=XX'+XX'=2XX' \in \nullsp(\mathcal A|_{\Sspace^n})$, and
\begin{equation}
    \ip{I_n}{\Delta} = 2\tr(XX') = 2\norm{X}^2 > 0 ,
\end{equation}
since $X$ has full column rank; \cref{thm:functional-id} then rules out identification.

On $\mathcal H_1$ it can be. Identifying a diagonal $\Delta = \diag(\delta)$ with $\delta \in \R^n$, \eqref{eq:diag-diag} shows that $\Proj_{\mathcal H_D} (M\Delta M)$ is represented by $(M \circ M) \delta$, so by \cref{thm:general-equivalence} $\nullsp(\mathcal A|_{\mathcal H_D}) = \{\diag(\delta):(M\circ M)\delta=0\}$, while $\ip{I_n}{\diag(\delta)}=\mathbf 1'\delta$. Hence $\phi_{I_n}$ is identified on $\mathcal H_D$ exactly when $\mathbf 1 \perp \nullsp(M\circ M)$, which holds in particular whenever $M \circ M$ is nonsingular, so by \cref{prop:leverage} it suffices that $\max_i h_{ii}<\tfrac12$.

The matched-pair design of \cref{ex:matched-pairs} gives an explicit instance, and shows that nonsingularity of $M\circ M$ is not needed. There $\nullsp(M \circ M) = \operatorname{span} \{(1,-1)'\}$, which is orthogonal to $\mathbf 1$, so $ \tr(\Sigma) = \sigma_1^2 + \sigma_2^2$ is identified on $\mathcal H_D$; indeed $\tr(\Sigma) = 2\,\E[\bu'\bu\mid X]$. On $\Sspace^n$, however, $\Sigma_0=I_2$ and $\Sigma_1 = \Sigma_0 + 2XX' = \left(\begin{smallmatrix} 3&2 \\ 2&3 \end{smallmatrix} \right)$ are both positive definite and satisfy $M \Sigma_0 M = M\Sigma_1M = M$, yet $\tr\Sigma_0 = 2\neq6 = \tr\Sigma_1$. The two covariance matrices are residual-second-moment equivalent, and $\Sigma_1$ is simply inadmissible under the diagonal model.
\end{proof}

\subsection{\texorpdfstring{Proof of \cref{lem:cone}}{Proof of cone}}

\begin{proof}
If $\Delta=0$ the claim is trivial, so assume $\Delta\neq0$ and set
\begin{equation}
    t:=\frac{\lambda_{\min}(\Sigma_0)}{\norm{\Delta}}>0,
\end{equation}
which is strictly positive because $\Sigma_0$ is positive definite.

\emph{Membership in $\mathcal H$.} Both $\Sigma_0$ and $\Delta$ lie in the linear space $\mathcal H$, hence so does $\Sigma_0 + s\Delta$ for every real $s$.

\emph{Positive semidefiniteness.} Let $v\in\R^n$ with $\|v\|=1$ and let $|s|\leq t$. By the variational characterization of the smallest eigenvalue and the Cauchy--Schwarz inequality,
\begin{align}
    v'(\Sigma_0+s\Delta)v
    &= v'\Sigma_0v+s\,v'\Delta v\\
    &\geq \lambda_{\min}(\Sigma_0)-|s|\,|v'\Delta v|\\
    &\geq \lambda_{\min}(\Sigma_0)-|s|\,\|\Delta\|_{\mathrm{op}}\\
    &\geq \lambda_{\min}(\Sigma_0)-t\,\norm{\Delta}
     = 0,
\end{align}
where the third line uses $|v'\Delta v|\leq\|\Delta\|_{\mathrm{op}}$ for unit $v$ and the fourth uses $\|\Delta\|_{\mathrm{op}}\leq\norm{\Delta}$, the operator norm being the largest singular value and the Frobenius norm the root of the sum of their squares. Hence $\Sigma_0 + s \Delta \in \Sspace^n_+$, and combining the two parts, $\Sigma_0 + s\Delta\in\mathcal H\cap\Sspace^n_+$ for all $|s|\leq t$.
\end{proof}

\subsection{\texorpdfstring{Proof of \cref{thm:functional-id}}{Proof of functional id}}

\begin{proof}
\emph{Step 1: (i) $\iff$ (ii).} Let $\Sigma_0,\Sigma_1$ be admissible and set $\Delta := \Sigma_1 - \Sigma_0 \in \mathcal H$. Then
\begin{equation}
    M\Sigma_1M = M\Sigma_0M \iff M(\Sigma_1-\Sigma_0)M=0 \iff \mathcal A(\Delta)=0 \iff \Delta \in \nullsp(\mathcal A),
\end{equation}
by linearity of $\mathcal A$. By bilinearity of the inner product the target values differ by
\begin{equation}
    \phi_R(\Sigma_1)-\phi_R(\Sigma_0)=\ip{R}{\Sigma_1}-\ip{R}{\Sigma_0}=\ip{R}{\Delta}.
\end{equation}
Thus $\phi_R$ takes equal values at every pair generating the same residual covariance precisely when $\ip{R}{\Delta}=0$ for all $\Delta \in \nullsp(\mathcal A)$. (By \cref{lem:cone} such a $\Delta$ is realized by an admissible pair at any positive definite $\Sigma_0 \in \mathcal H$, so the quantification is not vacuous.)

\emph{Step 2: (ii) $\iff$ (iii).} This is the definition of the orthogonal complement, taken inside $\mathcal H$; note $R\in\mathcal H$ and $\nullsp(\mathcal A) \subseteq \mathcal H$, so the complement is unambiguous.

\emph{Step 3: (iii) $\iff$ (iv).} We show $\range(\mathcal A^*)=\nullsp(\mathcal A)^\perp$. If $R=\mathcal A^*(D)$ for some $D$, then for every $\Delta\in\nullsp(\mathcal A)$,
\begin{equation}
    \ip{R}{\Delta}=\ip{\mathcal A^*(D)}{\Delta}=\ip{D}{\mathcal A(\Delta)}=\ip{D}{0}=0,
\end{equation}
so $\range(\mathcal A^*)\subseteq\nullsp(\mathcal A)^\perp$. For the reverse inclusion compare dimensions. Since $\rank(\mathcal A^*)=\rank(\mathcal A)$ and, by rank--nullity on $\mathcal H$, $\rank(\mathcal A)=\dim\mathcal H-\dim\nullsp(\mathcal A)$,
\begin{equation}
    \dim\range(\mathcal A^*) = \dim\mathcal H-\dim\nullsp(\mathcal A) = \dim\nullsp(\mathcal A)^\perp ,
\end{equation}
the last equality because both subspaces sit inside $\mathcal H$. A subspace contained in another of the same finite dimension equals it, which gives (iii) $\iff$ (iv).

\emph{Step 4: the adjoint formula.} For $\Delta\in\mathcal H$ and $D\in\Sspace^n$, using $M'=M$ and cyclicity of the trace,
\begin{align}
    \ip{\mathcal A(\Delta)}{D}
    &=\tr\big((M\Delta M)'D\big) =\tr(M\Delta'MD) =\tr(\Delta'MDM)\notag\\
    &=\ip{\Delta}{MDM} =\ip{\Delta}{\Proj_{\mathcal H}(MDM)},
\end{align}
the final step because $\Delta$ lies in $\mathcal H$ while $MDM-\Proj_{\mathcal H}(MDM)$ is orthogonal to it. By uniqueness of the adjoint, $\mathcal A^*(D)=\Proj_{\mathcal H}(MDM)$.
\end{proof}

\subsection{\texorpdfstring{Proof of \cref{cor:sandwich-nonid}}{Proof of sandwich nonid}}

\begin{proof}
\emph{Part (i).} Let $C=(X'X)^{-1}cc'(X'X)^{-1}$, which is symmetric, and $R_c=XCX'$. Taking $B=\tfrac12XC$ gives $XB'+BX' = \tfrac12XCX'+\tfrac12XCX' = XCX' = R_c$, so $R_c \in \nullsp(\mathcal A)$ by \cref{prop:kernel-exact}. It remains to check $R_c\neq0$. Since $X$ has full column rank and $c\neq0$, the vector $a=(X'X)^{-1}c$ is nonzero, hence $Xa\neq0$ and $R_c=(Xa)(Xa)'\neq0$. Therefore
\begin{equation}
    \ip{R_c}{R_c}=\norm{R_c}^2=\|Xa\|^4>0,
\end{equation}
so $R_c$ is a nonzero element of $\nullsp(\mathcal A)$ that is not orthogonal to itself; consequently $R_c\notin\nullsp(\mathcal A)^\perp$, and \cref{thm:functional-id} shows $\phi_{R_c}$ is not identified.

\emph{Part (ii).} From $HX=X$ we get $HR_cH=HXCX'H=XCX'=R_c$. Hence, for any $\Sigma$,
\begin{equation}
    \ip{R_c}{\Sigma}=\tr(R_c\Sigma)=\tr(HR_cH\Sigma)=\tr(R_c\,H\Sigma H),
\end{equation}
using cyclicity and $H'=H=H^2$, so $\phi_{R_c}$ depends on $\Sigma$ only through $H\Sigma H$. Moreover $R_c=XCX'$ with $C\in\Sspace^p$, so by \cref{prop:invariance}(iv) the matrix $R_c$ lies in the null space of the mean-free quadratic information map. Since $R_c\neq0$, that functional is not identified by the mean-free class either.
\end{proof}

\subsection{\texorpdfstring{Proof of \cref{prop:complete}}{Proof of complete}}

\begin{proof}
Write $A_F:=\mathcal A$ and $A_P:=P\mathcal A$, both with domain $\mathcal H$.

\emph{Step 0: the basic inclusion.} If $A_F\Delta=0$ then $A_P\Delta=P(A_F\Delta)=P0=0$, so
\begin{equation}\label{eq:app-kernel-incl}
    \nullsp(A_F)\subseteq\nullsp(A_P).
\end{equation}

\emph{Step 1: (i) $\implies$ (ii).} Assume $\nullsp(A_P)=\nullsp(A_F)$. By \cref{thm:functional-id} applied to each operator, the functionals identified from the full moments are those with $R \in \nullsp(A_F)^\perp$ and those identified from the selected moments are those with $R\in\nullsp(A_P)^\perp$. Equal null spaces give equal complements, so the two collections of identified functionals coincide; in particular every full-moment-identified functional is selected-moment identified.

\emph{Step 2: (ii) $\implies$ (i).} Assume every functional identified from the full
moments is identified from the selected moments, that is, $\nullsp(A_F)^\perp \subseteq \nullsp(A_P)^\perp$. Taking orthogonal complements inside $\mathcal H$ reverses the inclusion, giving $\nullsp(A_P)\subseteq\nullsp(A_F)$. Combined with \eqref{eq:app-kernel-incl} this yields equality.

\emph{Step 3: (i) $\iff$ (iii).} By Step 3 of the proof of \cref{thm:functional-id}, $\range(A^*) = \nullsp(A)^\perp$ for any linear $A$ on $\mathcal H$. Hence $\range(A_P^*) = \range(A_F^*)$ if and only if $\nullsp(A_P)^\perp = \nullsp(A_F)^\perp$, which by taking complements once more is equivalent to $\nullsp(A_P)=\nullsp(A_F)$.

\emph{Step 4: (i) $\iff$ (iv).} Rank--nullity on the common domain gives $\rank(A) = \dim \mathcal H-\dim\nullsp(A)$ for $A\in\{A_F,A_P\}$, so $\rank(A_P)=\rank(A_F)$ if and only if $\dim\nullsp(A_P)=\dim\nullsp(A_F)$. By \eqref{eq:app-kernel-incl} one null space contains the other, and nested subspaces of equal finite dimension coincide; the converse is immediate.
\end{proof}

\subsection{\texorpdfstring{Proof of \cref{lem:gram-general}}{Proof of gram general}}

\begin{proof}
\emph{Step 1: the adjoint.} Let $\Delta\in\mathcal H$ and $W\in\Sspace^n$. Using the definition of the inner product, then $M'=M$, then cyclicity of the trace,
\begin{align}
    \ip{\mathcal A(\Delta)}{W} &= \tr\big((M\Delta M)'W\big) = \tr(M'\Delta'M'W) = \tr(M\Delta'MW)\notag\\
                               &= \tr(\Delta'MWM) = \ip{\Delta}{MWM} = \ip{\Delta}{\Proj_{\mathcal H}(MWM)} ,
\end{align}
where the last equality holds because $MWM-\Proj_{\mathcal H}(MWM)$ is orthogonal to $\mathcal H$ and $\Delta \in \mathcal H$. Since this holds for all $\Delta$, uniqueness of the adjoint gives \eqref{eq:AF-adjoint}. Note that the transposes are carried throughout: the argument uses symmetry of $M$, not of $\Delta$.

\emph{Step 2: the Gram identity.} Apply Step 1 with $W=\mathcal A(\Delta)=M\Delta M$:
\begin{align}
    \mathcal A^*\mathcal A(\Delta) &= \Proj_{\mathcal H}\big(M(M\Delta M)M\big) = \Proj_{\mathcal H} \big(M^2\Delta M^2\big) = \Proj_{\mathcal H}(M\Delta M) = \mathcal A_S(\Delta),
\end{align}
using $M^2=M$ twice. As $\Delta\in\mathcal H$ was arbitrary, $\mathcal A_S=\mathcal A^*\mathcal A$, which is \eqref{eq:AS-gram}.

\emph{Step 3: self-adjointness and positivity.} For $\Delta_1,\Delta_2\in\mathcal H$,
\begin{equation}
    \ip{\mathcal A_S\Delta_1}{\Delta_2} =\ip{\mathcal A^*\mathcal A\Delta_1}{\Delta_2} =\ip{\mathcal A\Delta_1}{\mathcal A\Delta_2} =\ip{\Delta_1}{\mathcal A^*\mathcal A\Delta_2} =\ip{\Delta_1}{\mathcal A_S\Delta_2},
\end{equation}
so $\mathcal A_S$ is self-adjoint on $\mathcal H$. Taking $\Delta_1=\Delta_2=\Delta$ in the middle expression gives $\ip{\mathcal A_S\Delta}{\Delta}=\norm{\mathcal A\Delta}^2\geq0$, so $\mathcal A_S$ is positive semidefinite.
\end{proof}

\begin{remark}[Symmetry is not required]
The computation above uses only $M'=M=M^2$ and orthogonality of $\Proj_{\mathcal H}$; symmetry of $\Delta$ is never invoked, which is why the transposes are carried explicitly in the trace manipulations. The lemma therefore holds verbatim when $\mathcal H$ is a linear subspace of $\R^{n\times n}$ not contained in $\Sspace^n$, and so do \cref{thm:general-equivalence,thm:conditioning}, whose proofs appeal only to the Gram representation. This is the form in which the results are applied in \citet{chen2026robust}.
\end{remark}

\subsection{\texorpdfstring{Proof of \cref{thm:general-equivalence}}{Proof of general equivalence}}

\begin{proof}
\emph{Step 1: $\nullsp(\mathcal A)\subseteq\nullsp(\mathcal A_S)$.} If $\mathcal A\Delta=0$ then, by \cref{lem:gram-general}, $\mathcal A_S\Delta=\mathcal A^*\mathcal A\Delta=\mathcal A^*0=0$.

\emph{Step 2: $\nullsp(\mathcal A_S)\subseteq\nullsp(\mathcal A)$.} Suppose $\mathcal A_S\Delta=0$. Pairing with $\Delta$ and using the Gram identity together with the defining property of the adjoint,
\begin{equation}
    0=\ip{\Delta}{\mathcal A_S\Delta} =\ip{\Delta}{\mathcal A^*\mathcal A\Delta} =\ip{\mathcal A\Delta}{\mathcal A\Delta} =\norm{\mathcal A\Delta}^2 .
\end{equation}
A norm vanishes only at the zero element, so $\mathcal A\Delta=0$. Steps 1 and 2 give
\eqref{eq:general-kernel-equality}.

\emph{Step 3: the ranges.} By \cref{lem:gram-general} the operator $\mathcal A_S$ is self-adjoint, so $\mathcal A_S^*=\mathcal A_S$ and, by Step 3 of the proof of \cref{thm:functional-id} applied to $\mathcal A_S$, $\range(\mathcal A_S)=\range(\mathcal A_S^*)=\nullsp(\mathcal A_S)^\perp$. The same identity applied to $\mathcal A$ gives $\range(\mathcal A^*)=\nullsp(\mathcal A)^\perp$. Both complements are taken inside $\mathcal H$, and by Steps 1--2 the two null spaces coincide; hence
\begin{equation}
    \range(\mathcal A_S)=\nullsp(\mathcal A_S)^\perp =\nullsp(\mathcal A)^\perp=\range(\mathcal A^*),
\end{equation}
which is \eqref{eq:general-range-equality}.

\emph{Step 4: identification.} By \cref{thm:functional-id}, the functionals identified from the complete residual covariance are exactly those with $R\in\range(\mathcal A^*)$, and those identified from the structure-aligned information are exactly those with $R\in\range(\mathcal A_S^*)=\range(\mathcal A_S)$. Step 3 shows the two sets are equal.
\end{proof}

\subsection{\texorpdfstring{Proof of \cref{thm:conditioning}}{Proof of conditioning}}

\begin{proof}
\emph{Step 1: a singular value decomposition.} Let $m=\rank(\mathcal A)$ and write
\begin{equation}\label{eq:app-svd}
    \mathcal A=\sum_{j=1}^m\sigma_j\,w_j\ip{v_j}{\cdot}, \qquad \sigma_1\geq\cdots\geq\sigma_m>0,
\end{equation}
where $\{v_j\}_{j\le m}$ is orthonormal in $\mathcal H$ and spans $\nullsp(\mathcal A)^\perp$, and $\{w_j\}_{j\le m}$ is orthonormal in $\Sspace^n$. Such a decomposition exists for any linear map between finite-dimensional inner-product spaces.

\emph{Step 2: the spectrum of $\mathcal A_S$.} The adjoint of \eqref{eq:app-svd} is $\mathcal A^*=\sum_{j}\sigma_j\,v_j\ip{w_j}{\cdot}$. Composing and using $\ip{w_i}{w_j}=\delta_{ij}$,
\begin{equation}
    \mathcal A^*\mathcal A(\Delta) =\sum_{i=1}^m\sigma_i v_i\Big\langle w_i,\ \sum_{j=1}^m\sigma_jw_j\ip{v_j}{\Delta} \Big\rangle_F = \sum_{j=1}^m\sigma_j^2\,v_j\ip{v_j}{\Delta}.
\end{equation}
By \cref{lem:gram-general} the left side is $\mathcal A_S(\Delta)$, so $\mathcal A_S$ is diagonal in the orthonormal basis $\{v_j\}$ with eigenvalues $\sigma_j^2$, and it annihilates $\nullsp(\mathcal A)$. Its nonzero eigenvalues are therefore $\lambda_j=\sigma_j^2$ for $j=1,\ldots,m$, in the same order, which is
\eqref{eq:sv-squaring}.

\emph{Step 3: the condition numbers.} Since $\mathcal A_S$ is positive semidefinite by \cref{lem:gram-general}, its singular values equal its eigenvalues. Restricting both operators to $\nullsp(\mathcal A)^\perp=\operatorname{span}\{v_1,\ldots,v_m\}$, on which each is injective by \cref{thm:general-equivalence},
\begin{equation}
    \kappa(\mathcal A_S)=\frac{\lambda_1}{\lambda_m} = \frac{\sigma_1^2}{\sigma_m^2} = \left(\frac{\sigma_1}{\sigma_m}\right)^{\!2} = \kappa(\mathcal A)^2,
\end{equation}
which is \eqref{eq:cond-squaring}.

\emph{Step 4: the comparison.} By construction $\sigma_1\geq\sigma_m>0$, so $\kappa(\mathcal A)\geq1$ and therefore $\kappa(\mathcal A)^2\geq\kappa(\mathcal A)$, with equality if and only if $\kappa(\mathcal A)=1$, that is, if and only if all nonzero singular values of $\mathcal A$ coincide.
\end{proof}

\subsection{\texorpdfstring{Proof of \cref{cor:cluster-equivalence}}{Proof of cluster equivalence}}

\begin{proof}
\emph{Step 1: $P_C$ is the orthogonal projection onto $\mathcal H_C$.} Let $W\in\Sspace^n$ be partitioned conformably with the clusters as $W=(W_{gh})$. First, $P_C(W)=\diag(W_{11},\ldots,W_{GG})$ lies in $\mathcal H_C=\bigoplus_g\Sspace^{n_g}$, each diagonal block being symmetric because $W$ is. Second, the residual $W-P_C(W)$ has all diagonal blocks equal to zero, so for any $D=\diag(D_1,\ldots,D_G)\in\mathcal H_C$,
\begin{equation}
    \ip{D}{W-P_CW} = \sum_{g=1}^G\tr\Big(D_g'\big(W-P_CW\big)_{gg}\Big) = \sum_{g=1}^G\tr(D_g'\,0)=0 ,
\end{equation}
the first equality because the Frobenius inner product of a block-diagonal matrix with any matrix only involves the corresponding diagonal blocks. Thus $W-P_CW\perp\mathcal H_C$, and since $P_CW\in\mathcal H_C$, the decomposition $W=P_CW+(W-P_CW)$ identifies $P_C$ as $\Proj_{\mathcal H_C}$.

\emph{Step 2: apply the general theorem.} By Step 1, $P_C\mathcal A(\Delta)=\Proj_{\mathcal H_C}(M\Delta M)=\mathcal A_S(\Delta)$ for every $\Delta\in\mathcal H_C$, so $P_C\mathcal A$ is precisely the structure-aligned operator for $\mathcal H=\mathcal H_C$. \Cref{thm:general-equivalence} then gives $\nullsp(P_C\mathcal A)=\nullsp(\mathcal A)$, and the statement about identified functionals follows from \cref{thm:functional-id}.
\end{proof}

\subsection{\texorpdfstring{Proof of \cref{cor:no-cross-rescue}}{Proof of no cross rescue}}

\begin{proof}
Let $\Delta\in\mathcal H_C$ and suppose every within-cluster block of $M\Delta M$ vanishes, that is, $(M\Delta M)_{gg}=0$ for $g=1,\ldots,G$. By the definition of $P_C$ this says $P_C(M\Delta M)=0$, so $\Delta\in\nullsp(P_C\mathcal A)$. By \cref{cor:cluster-equivalence} the two null spaces coincide, so $\Delta\in\nullsp(\mathcal A)$, that is, $M\Delta M=0$. Every block of the zero matrix is zero, so in particular $(M\Delta M)_{gh}=0$ for all $g\neq h$. Hence no such $\Delta$ can have a nonzero cross-cluster block.
\end{proof}

\subsection{\texorpdfstring{Proof of \cref{prop:overid}}{Proof of overid}}

\begin{proof}
\emph{Step 1: definition.} For $Y\in\range(\mathcal A_S)$ choose some $\Delta\in\mathcal H$ with $\mathcal A_S\Delta=Y$, which exists by definition of the range, and set
\begin{equation}
    \mathcal T(Y):=(I-\Proj_{\mathcal H})(M\Delta M).
\end{equation}

\emph{Step 2: well-definedness.} Suppose $\Delta_1,\Delta_2\in\mathcal H$ both satisfy $\mathcal A_S \Delta_i = Y$. Then $\Delta_1 - \Delta_2 \in \nullsp(\mathcal A_S) = \nullsp(\mathcal A)$ by \cref{thm:general-equivalence}, so $M(\Delta_1-\Delta_2)M=0$ and therefore
\begin{equation}
    (I-\Proj_{\mathcal H})(M\Delta_1M)=(I-\Proj_{\mathcal H})(M\Delta_2M).
\end{equation}
The value of $\mathcal T(Y)$ does not depend on the choice of preimage.

\emph{Step 3: linearity.} Let $Y_1,Y_2\in\range(\mathcal A_S)$ with preimages $\Delta_1,\Delta_2$ and let $\alpha,\beta\in\R$. Then $\alpha\Delta_1+\beta\Delta_2$ is a preimage of $\alpha Y_1+\beta Y_2$, and since $\Delta\mapsto(I-\Proj_{\mathcal H})(M\Delta M)$ is linear, $\mathcal T(\alpha Y_1+\beta Y_2)=\alpha\mathcal T(Y_1)+\beta\mathcal T(Y_2)$.

\emph{Step 4: the identity.} For $\Sigma\in\mathcal H$, take $Y=\mathcal A_S\Sigma=\Proj_{\mathcal H}(M\Sigma M)$ with preimage $\Sigma$ itself. Then $\mathcal T(Y)=(I-\Proj_{\mathcal H})(M\Sigma M)$, which is \eqref{eq:overid}.
\end{proof}

\subsection{\texorpdfstring{Proof of \cref{lem:meanfree-invariant}}{Proof of meanfree invariant}}

\begin{proof}
\emph{Part (i).} Taking expectations in $y'Ay$ with $y=X\beta+\varepsilon$ and $\E[\varepsilon\mid X]=0$, $\Cov(\varepsilon\mid X)=\Sigma$,
\begin{equation}
    \E[y'Ay\mid X] =\beta'X'AX\beta+2\beta'X'A\,\E[\varepsilon\mid X]+\E[\varepsilon'A\varepsilon\mid X] = \beta'X'AX\beta+\tr(A\Sigma).
\end{equation}
The second term does not involve $\beta$, so the expectation is free of $\beta$ for every $\beta\in\R^p$ if and only if the quadratic form $\beta\mapsto\beta'X'AX\beta$ vanishes identically. As $X'AX$ is symmetric, this happens if and only if $X'AX=0$.

\emph{Part (ii).} Expanding for arbitrary $y\in\R^n$ and $b\in\R^p$,
\begin{equation}
    (y+Xb)'A(y+Xb)-y'Ay=2b'X'Ay+b'X'AXb .
\end{equation}
Fix $b$ and vary $y$: the right side is an affine function of $y$ whose linear part is $2(AXb)'y$, using $A'=A$. An affine function of $y$ vanishes identically only if its linear part does, so $AXb=0$; and then the constant term satisfies $b'X'AXb=(AXb)'Xb=0$ as well. Requiring this for every $b\in\R^p$ gives $AX=0$. Conversely, if $AX=0$ then both terms vanish for all $y$ and $b$.

\emph{Part (iii).} If $AX=0$ then $AH=AX(X'X)^{-1}X'=0$ and, transposing, $HA=(AH)'=0$ since $A$ and $H$ are symmetric. Hence
\begin{equation}
    A=(H+M)A(H+M)=HAH+HAM+MAH+MAM=MAM,
\end{equation}
because every term containing $H$ adjacent to $A$ vanishes. Conversely, if $A=MAM$ then $AX=MAMX=MA\,0=0$.
\end{proof}

\subsection{\texorpdfstring{Proof of \cref{prop:invariance}}{Proof of invariance}}

\begin{proof}
\emph{Part (i): dimensions.} Consider $\Psi:\Sspace^n\to\Sspace^p$, $\Psi(A)=X'AX$, which is linear. It is surjective: given $S\in\Sspace^p$, the matrix $A=X(X'X)^{-1}S(X'X)^{-1}X'$ is symmetric and satisfies
\begin{equation}
    \Psi(A)=X'X(X'X)^{-1}S(X'X)^{-1}X'X=S .
\end{equation}
Since $\Qclass=\nullsp(\Psi)$, rank--nullity gives
\begin{equation}
    \dim\Qclass=\dim\Sspace^n-\dim\Sspace^p=\frac{n(n+1)}{2}-\frac{p(p+1)}{2}.
\end{equation}
By \cref{lem:meanfree-invariant}(iii) the translation-invariant class is $\{MDM:D\in\Sspace^n\}$, which by Step 4 of the proof of \cref{prop:compression} is isomorphic to $\Sspace^r$ and so has dimension $r(r+1)/2$ with $r=n-p$. It is contained in $\Qclass$ because $X'MDMX=0$, and the containment is strict by the dimension count
\begin{equation}
    \dim\Qclass-\dim\{MDM\} =\frac{n(n+1)-p(p+1)-(n-p)(n-p+1)}{2} =p(n-p),
\end{equation}
which is positive since $p<n$. This is \eqref{eq:invariance-gap}.

\emph{Part (ii): the decomposition.} For any $A$, inserting $I_n=H+M$ twice gives $A=HAH+HAM+MAH+MAM$. If $A\in\Qclass$ then $X'AX=0$, and since
\begin{equation}
    HAH=X\big[(X'X)^{-1}X'AX(X'X)^{-1}\big]X',
\end{equation}
the bracketed matrix vanishes, so $HAH=0$. This yields
\eqref{eq:A-decomposition}.

\emph{Part (iii): what is identified.} Pairing \eqref{eq:A-decomposition} with $\Sigma$,
\begin{equation}
    \tr(A\Sigma)=\tr(MAM\Sigma)+\tr(HAM\Sigma)+\tr(MAH\Sigma).
\end{equation}
The first term equals $\tr\big(A(M\Sigma M)\big)$ by cyclicity and idempotence, hence is a linear functional of $M\Sigma M$. For the remaining two, put $N:=HAM$; then $MAH=(HAM)'=N'$, and since $\Sigma$ is symmetric,
\begin{equation}
    \tr(N\Sigma)+\tr(N'\Sigma)=2\tr(N\Sigma)=2\tr\big(A\,M\Sigma H\big),
\end{equation}
a linear functional of $M\Sigma H=(H\Sigma M)'$. Conversely every such functional arises: given any $N$ with $N=HNM$, the matrix $A:=N+N'$ is symmetric and satisfies $HAH=HNMH+HMN'H=0$, so $A\in\Qclass$ and $HAM=N$. Hence the mean-free class identifies all linear functionals of $M\Sigma M$ and of $H\Sigma M$, and no others.

\emph{Part (iv): the null space.} A perturbation $\Delta$ is invisible to the class exactly
when $\tr(A\Delta)=0$ for every $A\in\Qclass$, that is, $\Delta\in\Qclass^\perp$ within $\Sspace^n$. Since $\Qclass=\nullsp(\Psi)$ and $\Psi$ is surjective, $\Qclass^\perp=\range(\Psi^*)$. For $S\in\Sspace^p$ and $A\in\Sspace^n$,
\begin{equation}
    \ip{\Psi(A)}{S}=\tr\big((X'AX)'S\big)=\tr(A\,XSX')=\ip{A}{XSX'},
\end{equation}
so $\Psi^*(S)=XSX'$ and $\Qclass^\perp=\{XCX':C\in\Sspace^p\}$. The map $C\mapsto XCX'$ is injective because $X$ has full column rank, so this space has dimension $p(p+1)/2$, giving \eqref{eq:quad-kernel}. Finally, if $\phi_R$ depends on $\Sigma$ only through $H\Sigma H$ then $R=HRH$, hence $R=XCX'$ with $C=(X'X)^{-1}X'RX(X'X)^{-1}$, so $R\in\Qclass^\perp$; by the same argument as in \cref{thm:functional-id}, $\phi_R$ is identified only if $\ip{R}{R}=0$, that is, only if $R=0$.
\end{proof}

\subsection{\texorpdfstring{Proof of \cref{cor:leaveout-id}}{Proof of leaveout id}}

\begin{proof}
\emph{Step 1: $A_i$ is admissible and represents the leave-out moment.} With $A_i = \tfrac12 (b_ib_i'M + Mb_ib_i')$ we have $A_i'=A_i$, and since $MX=0$,
\begin{equation}
    X'A_iX=\tfrac12\big(X'b_ib_i'MX+X'Mb_ib_i'X\big)=0 ,
\end{equation}
so $A_i\in\Qclass$. Moreover, using $\bu=My$ and $y'Mb_ib_i'y=(b_i'My)(b_i'y)$,
\begin{equation}
    y'A_iy=\tfrac12\big[(b_i'y)(b_i'My)+(b_i'My)(b_i'y)\big]=y_i\bu_i .
\end{equation}
Taking expectations and using \cref{lem:meanfree-invariant}(i), $\E[y_i\bu_i\mid X]=\tr(A_i\Sigma)=(M\Sigma)_{ii}$.

\emph{Step 2: sufficiency of $h_{ii}<1$.} If $\Sigma$ is diagonal then $(M\Sigma)_{ii}=M_{ii}\sigma_i^2=(1-h_{ii})\sigma_i^2$. When $h_{ii}<1$ the factor $1-h_{ii}$ is nonzero, so
\begin{equation}
    \sigma_i^2=\frac{\E[y_i\bu_i\mid X]}{1-h_{ii}}
\end{equation}
is a known function of a mean-free quadratic expectation, hence identified.

\emph{Step 3: necessity.} Suppose $h_{ii}=1$, equivalently $M_{ii}=0$. Since $M$ is symmetric idempotent, $M_{ii}=b_i'Mb_i=b_i'M'Mb_i=\|Mb_i\|^2$, so $Mb_i=0$ and $b_i\in\col(X)$, say $b_i=Xd$. Then for every $A\in\Qclass$,
\begin{equation}
    \tr\big(A\,b_ib_i'\big)=b_i'Ab_i=d'X'AXd=0 .
\end{equation}
The perturbation $\Delta=b_ib_i'$ is diagonal, changes $\sigma_i^2$ by one unit, and is annihilated by every statistic in the class, so $\sigma_i^2$ is not identified. Combining the steps, the full diagonal is identified if and only if $\max_ih_{ii}<1$.
\end{proof}

\subsection{\texorpdfstring{Proof of \cref{rem:variance-cost}}{Proof of variance cost}}
\label{app:proof-variance-cost}

\begin{proof}
\emph{Step 1: the variance formula.} For $y\sim N(\mu,\Sigma)$ with $\mu=X\beta$ and symmetric $A$, the standard expression for the variance of a Gaussian quadratic form is
\begin{equation}
    \Var(y'Ay\mid X)=2\tr(A\Sigma A\Sigma)+4\mu'A\Sigma A\mu .
\end{equation}
Only the second term depends on $\beta$, and it equals $4\beta'\big(X'A\Sigma AX\big) \beta = 4 \| \Sigma^{1/2} AX \beta \|^2$ when $\Sigma\succ0$. This vanishes for every $\beta$ if and only if $AX=0$, that is, by \cref{lem:meanfree-invariant}(ii), if and only if $y'Ay$ is translation invariant.

\emph{Step 2: the leave-out statistic.} For $A_i$ in \eqref{eq:leaveout-A}, use $MX=0$:
\begin{equation}
    A_i\mu=\tfrac12\big(b_ib_i'MX\beta+Mb_ib_i'X\beta\big)
          =\tfrac12\,(b_i'X\beta)\,Mb_i
          =\tfrac12\,(x_i'\beta)\,Mb_i ,
\end{equation}
since $b_i'X=x_i'$ and $MX\beta=0$. Substituting into the $\beta$-dependent term,
\begin{equation}
    4\mu'A_i\Sigma A_i\mu = 4\cdot\tfrac14\,(x_i'\beta)^2\,b_i'M\Sigma Mb_i = (x_i'\beta)^2\,[M\Sigma M]_{ii},
\end{equation}
which is \eqref{eq:leaveout-variance}. The cost therefore grows with the squared fitted value at observation $i$.
\end{proof}

\subsection{\texorpdfstring{Proof of \cref{prop:dimension}}{Proof of dimension}}

\begin{proof}
By \cref{prop:compression}, $\mathcal A=\psi\circ\varphi$ where $\varphi:\Delta\mapsto Q'\Delta Q$ maps $\mathcal H$ into $\Sspace^r$ and $\psi:S\mapsto QSQ'$ is injective. Suppose $\nullsp(\mathcal A)=\{0\}$. If $\varphi(\Delta)=0$ for some $\Delta\in\mathcal H$ then $\mathcal A(\Delta)=\psi(0)=0$, so $\Delta=0$; hence $\varphi$ is injective on $\mathcal H$. An injective linear map cannot increase dimension, so
\begin{equation}
    \dim\mathcal H\leq\dim\varphi(\mathcal H)+\dim\nullsp(\varphi) = \dim\varphi(\mathcal H) \leq \dim \Sspace^r = \frac{r(r+1)}{2}. \qedhere
\end{equation}
\end{proof}

\subsection{\texorpdfstring{Proof of \cref{prop:leverage}}{Proof of leverage}}

\begin{proof}
Identify a diagonal $\Delta=\diag(\delta)$ with the vector $\delta\in\R^n$.

\emph{Part (i).} By \eqref{eq:diag-diag} the diagonal of $M\Delta M$ is $(M\circ M)\delta$, and since $\Proj_{\mathcal H_D}$ extracts the diagonal, the structure-aligned operator is represented by the matrix $M\circ M$. By \cref{thm:general-equivalence} its null space coincides with $\nullsp(\mathcal A)$, so $\nullsp(\mathcal A)=\{0\}$ if and only if $(M\circ M)\delta=0$ implies $\delta=0$, that is, if and only if $M\circ M$ is nonsingular.

\emph{Part (ii).} Suppose $h_{ii}=1$, so $M_{ii}=1-h_{ii}=0$. Using $M'=M=M^2$,
\begin{equation}
    M_{ii}=b_i'Mb_i=b_i'M'Mb_i=\|Mb_i\|^2 ,
\end{equation}
hence $Mb_i=0$. Therefore $\mathcal A(b_ib_i')=Mb_ib_i'M=(Mb_i)(Mb_i)'=0$, so the nonzero diagonal direction $b_ib_i'$ lies in $\nullsp(\mathcal A)$. Taking $R=b_ib_i'$ gives $\phi_R(\Sigma)=\sigma_i^2$ and $\ip{R}{b_ib_i'}=1\neq0$, so by \cref{thm:functional-id} the variance $\sigma_i^2$ is not identified.

\emph{Part (iii).} The matrix $M\circ M$ is symmetric with entries $M_{ij}^2\geq0$. From $M=M^2$ and $M'=M$ we get $M_{ii}=\sum_jM_{ij}^2$, so the $i$th row of $M\circ M$ has diagonal entry $M_{ii}^2=(1-h_{ii})^2$ and off-diagonal row sum
\begin{equation}
    \sum_{j\neq i}M_{ij}^2=M_{ii}-M_{ii}^2=(1-h_{ii})h_{ii}.
\end{equation}
If $h_{ii}<\tfrac12$ then $1-h_{ii}>h_{ii}$, and multiplying by $1-h_{ii}>0$ gives $(1-h_{ii})^2>(1-h_{ii})h_{ii}$, so row $i$ is strictly diagonally dominant. If this holds for every $i$, then $M\circ M$ is a symmetric, strictly diagonally dominant matrix with positive diagonal entries, hence positive definite by Gershgorin's theorem and in particular nonsingular. By Part (i) the full diagonal of $\Sigma$ is then identified.
\end{proof}

\section{Additional algebra}\label{app:algebra}

\subsection{Dimension decomposition}

Let $r=n-p$. The symmetric covariance space has dimension $n(n+1)/2$. The three blocks associated with the orthogonal decomposition $\R^n = \col(X) \oplus \col(X)^\perp$ have dimensions
\begin{equation}
    \dim(M\Sigma M) = \frac{r(r+1)}{2}, \qquad \dim(H\Sigma M) = pr, \qquad \dim(H\Sigma H) = \frac{p(p+1)}{2},
\end{equation}
which add to $n(n+1)/2$. Residual second moments access the first block. Mean-free quadratic statistics access the first two. The fitted-space block remains inaccessible without further restrictions, and by \cref{cor:sandwich-nonid} it is the block on which the sandwich variance depends.

\subsection{Numerical diagnostic for alternative moment selections}
\label{app:diagnostic}

Let $\{E_1,\ldots,E_m\}$ be an orthonormal basis of $\mathcal H$. Construct the matrix representation of the full residual-information operator using columns
\begin{equation}
    \operatorname{vec}(ME_jM), \qquad j=1,\ldots,m.
\end{equation}
For a proposed selection $P$, construct the corresponding columns $\operatorname{vec}( P(ME_jM) )$. The selected moment collection is information complete if and only if $\rank(P\mathcal A) = \rank(\mathcal A)$.

In numerical applications, singular values rather than literal floating-point ranks should be examined. By \cref{thm:conditioning}, the structure-aligned selection $P=\Proj_{\mathcal H}$ always attains rank equality but returns squared singular values; a rank computed by thresholding at a fixed tolerance can therefore appear to fail for the aligned system while succeeding for the full one. Comparisons should be made on a common scale, for instance by comparing $\{\lambda_j(\mathcal A_S)^{1/2}\}$ with $\{\sigma_j(\mathcal A)\}$.

For a block-diagonal covariance space $\mathcal H_C=\bigoplus_g\Sspace^{n_g}$ the operator has an explicit matrix form. With $E_g\in\R^{n\times n_g}$ the selection matrix of cluster $g$, $N_g:=ME_g$, and $D_{n_g}$ the duplication matrix satisfying $\operatorname{vec}(S)=D_{n_g}\operatorname{vech}(S)$ for $S\in\Sspace^{n_g}$,
\begin{equation}\label{eq:app-cluster-KR}
    \operatorname{vec}(M\Delta M) = \mathcal K\begin{pmatrix} \operatorname{vech}(\Delta_1)\\ \vdots\\ \operatorname{vech}(\Delta_G) \end{pmatrix},
    \qquad
    \mathcal K := \big[\,(N_1\otimes N_1)D_{n_1},\ \ldots,\ (N_G\otimes N_G)D_{n_G}\,\big] ,
\end{equation}
so $\nullsp(\mathcal A|_{\mathcal H_C})=\{0\}$ if and only if $\mathcal K$ has full column rank $\sum_gn_g(n_g+1)/2$. The matrix $[\,N_g\otimes N_g\,]_{g=1}^G$ is the Khatri--Rao product of $M$ with itself relative to the cluster partition of its columns, and $\mathcal K$ is its restriction to symmetric blocks; the restriction is not cosmetic, since $N_gKN_g'$ is antisymmetric whenever $K$ is and the symmetric and antisymmetric images are orthogonal, so full column rank of the unrestricted product is strictly stronger than \eqref{eq:app-cluster-KR} requires. The Gram matrix of $\mathcal K$ has $(g,h)$ block $D_{n_g}'(M_{gh} \otimes M_{gh})D_{n_h}$, equivalently the pairing $\ip{N_gSN_g'}{N_hTN_h'}=\tr(SM_{gh}TM_{hg})$ on $\Sspace^{n_g} \times \Sspace^{n_h}$. When every $n_g=1$ the duplication matrices are scalars, \eqref{eq:app-cluster-KR} reduces to the columnwise Khatri--Rao product $M\odot M$, and its Gram is $(M'M)\circ(M'M)=M\circ M$ by idempotence, recovering \cref{prop:leverage}(i). For a general transformation $T$ the same computation returns $(T'T)\circ(T'T)$ rather than $T\circ T$, which is \cref{rem:scope-TT} in coordinates.

For the mean-free quadratic information class, choose a basis $\{A_1,\ldots,A_q\}$ of $\Qclass = \nullsp(A\mapsto X'AX)$. The matrix with entries $\tr(A_kE_j)$ maps latent covariance directions into mean-free quadratic expectations. Its rank gives the dimension of the covariance-functional space identified by the larger information class.

\subsection{The computation behind \texorpdfstring{\cref{ex:cluster-fe}}{the worked example}}
\label{app:worked-svd}

With $X=[\mathbf 1,v]$ and $v=(1,0,0,-1)'$ we have $X'X=\diag(4,2)$ and $H=\tfrac14\mathbf 1\mathbf 1' + \tfrac12 vv'$, so
\begin{equation}\label{eq:app-worked-M}
    M=\frac14\begin{pmatrix} 1&-1&-1&1\\ -1&3&-1&-1\\ -1&-1&3&-1\\ 1&-1&-1&1 \end{pmatrix} .
\end{equation}
Throughout, $\mathcal H_C$ carries the orthonormal coordinates $(a_1,\sqrt2c_1,d_1,a_2,\sqrt2c_2,d_2)$.

\emph{The full map.} The residual subspace
$\col(X)^\perp=\{w:w'\mathbf 1=0,\ w_1=w_4\}$ has orthonormal basis
\begin{equation}
    u_1=\tfrac1{\sqrt2}(0,1,-1,0)', \qquad u_2=\tfrac12(1,-1,-1,1)' ,
\end{equation}
and $Q=[u_1,u_2]$ satisfies $QQ'=M$. By \cref{prop:compression}, $\mathcal A=\psi\circ\varphi$ with $\varphi(\Delta)=Q'\Delta Q$ and $\psi(S)=QSQ'$; since $\|QSQ'\|_F^2=\tr(S^2Q'Q)=\|S\|_F^2$, the map $\psi$ is an isometry, so the singular values of $\mathcal A$ are those of $\varphi:\mathcal H_C\to\Sspace^2$ and $\rank(\mathcal A) \leq \dim \Sspace^2 = 3$, which is \cref{prop:dimension} here. Restricting $u_1$ to the two clusters gives $(0,\tfrac1{\sqrt2})$ and $(-\tfrac1{\sqrt2},0)$, and $u_2$ gives $\tfrac12(1,-1)$ and $\tfrac12(-1,1)$, so for block-diagonal $\Delta$,
\begin{equation}
    S_{11}=\tfrac12(d_1+a_2), \qquad
    S_{22}=\tfrac14\sum_{g}\left(a_g-2c_g+d_g\right), \qquad
    S_{12}=\tfrac1{2\sqrt2}\left(c_1-d_1+a_2-c_2\right).
\end{equation}
In the coordinates above on $\mathcal H_C$ and $(S_{11},S_{22},\sqrt2S_{12})$ on $\Sspace^2$, the map $\varphi$ is represented by
\begin{equation}\label{eq:app-worked-G}
    G=\begin{pmatrix}
        0 & 0 & \tfrac12 & \tfrac12 & 0 & 0\\[2pt]
        \tfrac14 & -\tfrac{\sqrt2}4 & \tfrac14 & \tfrac14 & -\tfrac{\sqrt2}4 & \tfrac14\\[2pt]
        0 & \tfrac1{2\sqrt2} & -\tfrac12 & \tfrac12 & -\tfrac1{2\sqrt2} & 0
    \end{pmatrix},
    \qquad
    GG'=\begin{pmatrix} \tfrac12 & \tfrac14 & 0\\ \tfrac14 & \tfrac12 & 0\\ 0 & 0 & \tfrac34 \end{pmatrix} .
\end{equation}
The eigenvalues of $GG'$ are $\tfrac12\pm\tfrac14$ from the leading block together with $\tfrac34$, so
\begin{equation}\label{eq:app-worked-sv}
    \sigma(\mathcal A)=\left(\tfrac{\sqrt3}2,\ \tfrac{\sqrt3}2,\ \tfrac12\right), \qquad \rank(\mathcal A)=3, \qquad \dim\nullsp(\mathcal A)=3 .
\end{equation}

\emph{The aligned map.} \Cref{thm:general-equivalence} and \cref{thm:conditioning} can
be verified directly rather than read off the Gram identity. The operator $\mathcal A_S (\Delta) = \Proj_{\mathcal H_C}(M\Delta M)$ retains the two diagonal blocks of $M\Delta M$, which are most compactly recorded through the compression: since $M\Delta M = Q\varphi(\Delta)Q'$, each aligned entry is a fixed linear combination of $(S_{11},S_{22},S_{12})$, and pairing two aligned images gives
\begin{equation}
    \ip{\mathcal A_S\Delta}{\Delta'}=\ip{M\Delta M}{\Delta'} =\ip{\varphi(\Delta)}{\varphi(\Delta')} ,
\end{equation}
the first equality because $\Delta'\in\mathcal H_C$ and the second by the isometry of $\psi$. Hence $\mathcal A_S$ is represented by $G'G$, a $6\times6$ positive semidefinite matrix of rank three whose nonzero eigenvalues are those of $GG'$ in \eqref{eq:app-worked-G}, namely $(\tfrac34,\tfrac34,\tfrac14)$ --- exactly the squares of \eqref{eq:app-worked-sv}. The null spaces agree in dimension, as \cref{thm:general-equivalence} requires, and
\begin{equation}
    \kappa(\mathcal A)=\frac{\sqrt3/2}{1/2}=\sqrt3, \qquad \kappa(\mathcal A_S)=\frac{3/4}{1/4}=3=\kappa(\mathcal A)^2 .
\end{equation}
The rank attains the bound of \cref{prop:dimension} exactly, so the loss of three covariance directions in \cref{ex:cluster-fe} is forced by counting alone.
\end{document}